\documentclass[prb,onecolumn,showpacs,preprintnumbers,amsmath,amssymb]{revtex4}
\usepackage{graphicx}
\usepackage{setspace}
\usepackage{bm}
\begin{document}
\begin{spacing}{1.3}
\pagenumbering{arabic}
\title{Spectrum of electronic excitations due to the adsorption of atoms on metal surfaces}

\author{M. S. Mizielinski}
\affiliation{Department of Physics, University of Bath, Bath BA2 7AY, UK}
\author{D. M. Bird}
\affiliation{Department of Physics, University of Bath, Bath BA2 7AY, UK}
\author{M. Persson}
\affiliation{Department of Chemistry, University of Liverpool, Liverpool L69 3BX, UK \bigskip}
\author{S. Holloway}
\affiliation{Department of Chemistry, University of Liverpool, Liverpool L69 3BX, UK \bigskip}
\begin{abstract}
The time-dependent, mean-field Newns-Anderson model for a spin-polarised adsorbate
approaching a metallic surface is solved in the wide-band limit. Equations for the
time-evolution of the electronic structure of the adsorbate-metal system are derived
and the spectrum of electronic excitations is found. The behaviour of the model
is demonstrated for a set of physically reasonable parameters.
\end{abstract}
\pacs{73.20.Hb, 34.50.Dy, 68.43.-h}
\maketitle

\newcommand{\vak}{V_{ak}}
\newcommand{\vakp}{V_{ak'}}
\newcommand{\vakpp}{V_{ak''}}
\newcommand{\ebar}{\bar{\epsilon}_{a\sigma}}
\newcommand{\ebarad}{\bar{\epsilon}_{a\sigma}^{(ad)}}
\newcommand{\eks}{\epsilon_{k\sigma}}
\newcommand{\eksp}{\epsilon_{k'\sigma}}
\newcommand{\ekspp}{\epsilon_{k''\sigma}}
\newcommand{\etilde}{\tilde{\epsilon}_{a\sigma}}
\newcommand{\ns}{n_{\sigma}}
\newcommand{\nas}{n_{a\sigma}}
\newcommand{\nasdot}{\dot{n}_{a\sigma}}
\newcommand{\nkks}{n_{kk\sigma}}
\newcommand{\nkkps}{n_{kk'\sigma}}
\newcommand{\qs}{q_{\sigma}}
\newcommand{\rhoinst}{\rho_{a\sigma}^{(inst)}}
\newcommand{\Pinst}{p^{(inst)}_{\sigma}}
\newcommand{\gpi}[1]{\sqrt{\frac{\Gamma(#1)}{2\pi}}}

\newcommand{\nams}{n_{a-\sigma}}
\newcommand{\ca}{\hat c^{\phantom{\dagger}}_{a\sigma}}
\newcommand{\ck}{\hat c^{\phantom{\dagger}}_{k\sigma}}
\newcommand{\ckp}{\hat c^{\phantom{\dagger}}_{k'\sigma}}
\newcommand{\cad}{\hat c^{\dagger}_{a\sigma}}
\newcommand{\ckd}{\hat c^{\dagger}_{k\sigma}}
\newcommand{\ckpd}{\hat c^{\dagger}_{k'\sigma}}
\newcommand{\ps}{p_{\sigma}}
\newcommand{\naks}{n_{ak\sigma}}
\newcommand{\nkas}{n_{ka\sigma}}
\newcommand{\nbbps}{n_{bb'\sigma}}
\newcommand{\rs}{r_{\sigma}}
\newcommand{\bra}[1]{\langle #1 \vert}
\newcommand{\ket}[1]{\vert #1 \rangle}
\newcommand{\braket}[2]{\langle #1 \vert #2 \rangle}
\newcommand{\as}{a_{\sigma}}
\newcommand{\ks}{k_{\sigma}}
\newcommand{\ksp}{k_{\prime\sigma}}
\newcommand{\mus}{\mu_\sigma}
\newcommand{\nust}{\nu_{\sigma t}}
\newcommand{\musp}{\mu_{\prime\sigma}}
\newcommand{\epsmutz}{\epsilon^\mu_{\sigma t_0}}
\newcommand{\epsnut}{\epsilon^\nu_{\sigma t}}
\renewcommand{\Re}{\textrm{Re}}
\renewcommand{\Im}{\textrm{Im}}
\newcommand{\nsi}{n_\sigma^{(1)}}
\newcommand{\nsii}{n_\sigma^{(2)}}
\newcommand{\nsiii}{n_\sigma^{(3)}}
\newcommand{\nsiv}{n_\sigma^{(4)}}
\newcommand{\NA}{Newns-Anderson}
\newcommand{\Gs}{G^\sigma}
\newcommand{\GsI}{G^{0,\sigma}}
\newcommand{\Gaa}{G^{\sigma}_{aa}}
\newcommand{\Gak}{G^{\sigma}_{ak}}
\newcommand{\Gka}{G^{\sigma}_{ka}}
\newcommand{\Gkk}{G^{\sigma}_{kk}}
\newcommand{\Gkkp}{G^{\sigma}_{kk'}}
\newcommand{\Gkpk}{G^{\sigma}_{k'k}}
\newcommand{\GaaI}{G^{0\sigma}_{aa}}
\newcommand{\GkkI}{G^{0\sigma}_{kk}}
\newcommand{\GkpkpI}{G^{0\sigma}_{k'k'}}
\newcommand{\Gakp}{G^{\sigma}_{ak'}}
\newcommand{\Gbbps}{G^{\sigma}_{bb'}}
\newcommand{\hs}{\hat{h}_{\sigma}}
\newcommand{\hsone}{\hat{h}_{\sigma}^{(1)}}
\newcommand{\cs}{c_{\sigma}}
\newcommand{\bs}{b_{\sigma}}
\newcommand{\bps}{b'_{\sigma}}

\section{Introduction}
\label{sec:intro}

Great strides have been made in recent years in our understanding of the adsorption and reaction of molecules on surfaces.
Much of this progress is due to the development of electronic structure methods based on density functional theory (DFT),
from which the potential energy surface (PES) for the molecule-surface interaction can be obtained. When combined with classical
or quantum dynamical calculations, the ab-initio PES provides a clear picture of the molecular pathways and the making and breaking
of bonds\cite{darling95}. It is well known that DFT-based calculations are approximate in their use of a model for the exchange-correlation
potential. However, for any dynamical process there is another key approximation made. DFT calculations are firmly rooted in
the Born-Oppenheimer approximation (BOA); the electronic system remains in the ground state throughout the molecule-surface encounter
and all the dynamics takes place on the adiabatic PES\cite{wodtke04}. However, the BOA cannot strictly be valid for interactions
involving a metallic substrate. This is because there is a continuum of electronic states at the Fermi level, and consequently any
dynamical process will lead to the excitation of electron-hole pairs\cite{wodtke04}.

How large is the coupling between the nuclear motion and the electronic system? Some very recent studies have suggested that non-adiabatic
effects are small for closed shell molecules. Dynamical calculations for H$_2$/Pt\cite{nieto06} and N$_2$/Ru\cite{diaz06} based on
the ground state PES show good agreement with molecular beam scattering experiments, suggesting that coupling to electron-holes
pairs is not significant for these
systems. However, there is strong evidence in other cases that non-adiabatic effects are large. Two recent series of experiments have
provided striking evidence of electronic excitations in surface reactions. First, Nienhaus and co-workers
\cite{nienhaus99,gergen01,nienhaus02a,nienhaus02b} adsorbed a range of atomic and molecular species on thin metal films of silver and
copper which made up one contact of a Schottky diode. The hot electrons or holes resulting from the dissipation of the chemisorption
energy were detected as a chemically induced current, or chemicurrent. Second, White and co-workers\cite{white05,white06} investigated the
scattering of vibrationally excited NO molecules from a low work function, caesium-doped gold surface. For molecular beams with
a vibrational energy greater than the work function the emission of exo-electrons was observed, showing that electronic excitation
is a significant channel for the dissipation of vibrational energy.

Most theoretical treatments of non-adiabatic effects use a ``nearly-adiabatic'' approximation in which the perturbation of the
electronic system is assumed to be weak and slow. This leads to a friction-based description of the energy transfer between the
adsorbate and the
substrate (see \cite{trail03} and references therein). The friction coefficient can be calculated using ab-initio methods\cite{trail01},
and has been applied to vibrational damping\cite{persson04} and desorption dynamics\cite{luntz05}. By making a connection between the
friction description and the forced oscillator model it is also possible to obtain the spectrum of electron-hole pair
excitations\cite{trail02,trail03}. However, the friction-based description is not sufficient for strongly non-adiabatic effects, and
it fails completely for the adsorption of spin-polarised species. Trail et al.\cite{trail02,trail03} attempted to use the nearly-adiabatic
approximation to model the excitation of chemicurrents in the adsorption of H atoms on Cu. They found that the friction coefficient
diverges at the point in the trajectory where the ground state has a transition from being spin polarised (H-atom far from the surface)
to non-polarised (H close to the surface). Non-adiabatic effects in this case are large and a theoretical description must go beyond
the friction approximation.

In previous work\cite{bird04,mizielinski05} we have attempted to understand the origin of this divergence in the friction coefficient
by modelling a strongly non-adiabatic coupling between an adsorbate and a metal surface. We used the simplest possible model of a
gas-surface interaction: the Newns-Anderson model\cite{anderson61,newns69}, where a single adsorbate level interacts with a wide
electronic band. Within the mean-field approximation we showed that the time dependence of the occupancy of the adsorbate state can be
found analytically, and we derived expressions for the rate of energy transfer to the surface\cite{mizielinski05}. The Newns-Anderson
model shows the same ground-state spin transition as the ab-initio DFT calculations, but the singularity in the energy transfer is
removed in the fully non-adiabatic solution. In our previous work we did not solve for the evolution of the substrate states, and
so we could not derive the spectrum of electronic excitations which would be required, for example, in modelling chemicurrents.
Our aim in this paper is therefore to extend our previous work to a full solution of the non-adiabatic dynamics. In section
\ref{sec:dyn_spec} we derive an expression for the distribution of occupied electronic states and how this evolves with time. The
equivalent distribution without electronic excitation is considered in section \ref{sec:ad_spec}, which allows us to calculate the
spectrum of excitations. Section \ref{sec:computation} gives an outline of the methodology we have used to compute the excitation
spectrum. We demonstrate the behaviour of the model in section \ref{sec:num_res}, and concluding remarks are made in
section \ref{sec:conclusion}.

\section{Distribution of occupied states in the Newns-Anderson model}
\label{sec:dyn_spec}

Our expression for the distribution of electronic states is
obtained from the one-electron density matrix and is derived
within the framework of the time-dependent, mean-field
\NA\cite{anderson61,newns69} model. This is defined by the
Hamiltonian
\begin{subequations}
  \begin{eqnarray}
    \hat{H}(t) &=& \sum_\sigma \hat{H}_\sigma(t) - U\nas(t)\nams(t),
    \label{eq:many_e_ham}
  \end{eqnarray}
with
  \begin{eqnarray}
    \hat{H}_\sigma(t) &=& \ebar(t) \cad \ca + \sum_{k} \eks \ckd \ck
                + \sum_{k} \left(\vak(t)\cad \ck  + \textrm{H.c.}\right)
    \label{eq:single_spin_many_e_ham}
  \end{eqnarray}
\end{subequations}
where $\hat{H}(t)$ represents the total energy of the system and $\hat{H}_\sigma(t)$ is the many-electron Hamiltonian for spin $\sigma$.
$\cad$ and $\ckd$ are the creation operators for electrons in the adsorbate, $\ket{\as}$, and metal, $\ket{\ks}$,
states, respectively. $\vak(t)$ is the interaction potential, $U$ is the intra-adsorbate Coulomb repulsion energy
and $\eks$ is the energy of the metal state $\ket{\ks}$. $\ebar$ is the mean-field energy level of the adsorbate state and
is defined as
\begin{equation}
  \ebar(t) = \epsilon_a(t) + U\nams(t),
  \label{eq:ebar_defn}
\end{equation}
where the time-dependent occupation of the $\ket{\as}$ state
\begin{equation}
  \nas(t) = \bra{a_\sigma }\hat{n}_{1\sigma}(t)\ket{a_\sigma}
  \label{eq:mats_nas}
\end{equation}
is determined by the one-electron density matrix
\begin{equation}
  \bra{\bs^\prime}\hat{n}_{1\sigma}(t)\ket{\bs} \equiv
  \langle\!\langle \hat{c}_{b\sigma}^\dagger(t)\hat{c}_{b^\prime\sigma}(t) \rangle\!\rangle .
  \label{eq:mats_onematrix}
\end{equation}
Here $\ket{b_\sigma}$ refer to one of the basis states
$\ket{a_\sigma}$ and $\ket{k_\sigma}$, $\langle\!\langle . \rangle\!\rangle$ denotes a thermal
average and the annihilation and creation operators are given in
the Heisenberg picture. As in our previous
paper\cite{mizielinski05} we assume that the evolution of the
system is driven by the variation in the bare adsorbate energy
level, $\epsilon_a(t)$, and the interaction potential, $\vak(t)$. The
Coulomb repulsion energy $U$ is assumed to be constant. In order
to model the behaviour of a real system, the variation of these
parameters can be estimated from DFT calculations, as discussed in
our previous work\cite{mizielinski05}.

Since the mean-field Hamiltonian is quadratic in the annihilation
and creation operators, the time-evolution of the one-electron
density matrix, $\hat{n}_{1\sigma}(t)$, can be obtained by considering
the time-evolution of the one-electron states of the one-electron
Hamiltonian,
\begin{equation}
  \hs(t) = \ebar(t)\ket{\as}\bra{\as} + \sum_k \eks\ket{\ks}\bra{\ks}
  + \sum_k \left(\vak(t)\ket{\as}\bra{\ks}+\textrm{H.c.}\right).
\label{eq:schro_ham_one}
\end{equation}
We will use $\ket{\mus(t)}$ to represent the time-evolving set of
electronic states of the one-electron Hamiltonian $\hs$. These
will evolve from an initial state at time $t_0$ which is one of
the basis states $\ket{\bs}$ of the system (i.e. $\ket{\as}$ or
$\ket{\ks}$). We note here that the initial interaction potential
must be zero (i.e. $\vak(t_0)=0$). The time dependence of the
creation and annihilation operators is now simply determined by
the time dependence of the one-electron states as
\begin{equation}
  \hat{c}_{b\sigma}(t)=\sum_{b^\prime}\braket{\bs}{\mus^\prime(t)}\hat{c}_{b^\prime\sigma}(t_0)
\label{eq:op_relation}
\end{equation}
where $\ket{\mus^\prime(t_0)} = \ket{b^\prime}$. This result and
(\ref{eq:mats_onematrix}) shows that $\hat{n}_{1\sigma}(t)$ can
be expressed in terms of $\ket{\mus(t)}$ and the initial occupation
$f(\epsmutz)$ of the dynamical state $\ket{\mus(t)}$ at time
$t_0$, where $f$ is the Fermi function, as
\begin{equation}
\hat{n}_{1\sigma}(t) = \sum_\mu
\ket{\mus(t)}\bra{\mus(t)}f(\epsmutz) .
\label{eq:onematrixresult}
\end{equation}

The quantity we are interested in is the distribution of occupied
one-electronic states and how this evolves with time. In order to
derive an expression for this distribution function, we need to
take the diagonal matrix elements of $\hat{n}_{1\sigma}(t)$ with
respect to eigenstates of $\hs$. We label these eigenstates as
$\ket{\nust}$ with energies $\epsnut$, and the distribution
function becomes
\begin{eqnarray}
  \ns(\epsilon,t) &\equiv & \sum_\nu \bra{\nust}\hat{n}_{1\sigma}\ket{\nust} \delta(\epsilon-\epsnut), \nonumber \\
  & =& \sum_{\mu,\nu} \vert \langle \nust \vert \mus(t) \rangle \vert^2 f(\epsmutz) \delta(\epsilon-\epsnut).
  \label{eq:elec_spec_dyn}
\end{eqnarray}
It is important to note that we choose states $\ket{\nust}$ that
are not the usual eigenstates of the Newns-Anderson model. In the
static case $\epsilon_a(t)$ and $\vak(t)$ are held constant at a
given value and \eqref{eq:single_spin_many_e_ham} and \eqref{eq:ebar_defn} 
are solved self-consistently, which means that $\nas$ and $\ebar$
are determined entirely by these parameters. This is what we refer
to as the adiabatic state of the system. In our system, however,
$\nas$ is not the self-consistent solution, but is the occupation
of the adsorbate orbital at a given instant and therefore depends
on how the system has evolved. The states $\ket{\nust}$ are the
instantaneous states of the one-electron Hamiltonian rather than
the adiabatic states. This choice of eigenstate means that the
total energy of the system is given by the first moment of the
distribution function, that is
\begin{subequations}
\begin{eqnarray}
  E(t) &=& \sum_\sigma \int d\epsilon\, \epsilon \ns(\epsilon,t) - U \nas(t)\nams(t)
  \label{eq:elec_spec_dyn_energy_1} \\
  &=& \sum_{\sigma,\mu} \bra{\mus(t)}\hs(t)\ket{\mus(t)}f(\epsmutz) - U \nas(t)\nams(t)
  \label{eq:elec_spec_dyn_energy_2} \\
  &=& \sum_{\sigma} \langle\!\langle \hat{H}_\sigma(t) \rangle\!\rangle - U \nas(t)\nams(t).
  \label{eq:elec_spec_dyn_energy_3}
\end{eqnarray}
\end{subequations}
This relationship between the distribution function and the total energy provides a useful check on the result for
$\ns(\epsilon,t)$ presented later. We also note that in the long time limit, when evolution of the system is finished,
the eigenstates $\ket{\nust}$ converge to the adiabatic states.

In order to express the distribution function in a more manageable form we use Green's functions
to replace the instantaneous states $\ket{\nust}$.
We rewrite (\ref{eq:elec_spec_dyn}) as
\begin{equation}
  \ns(\epsilon,t) = -\frac{1}{\pi} \textrm{Im} \left\{{\rm Tr}[\hat{n}_{1\sigma}(t)\Gs(\epsilon;t)]\right\},
  \label{eq:elec_spec_dyn_2}
\end{equation}
where $\Gs$ is the instantaneous Green's function defined as
\begin{equation}
  \Gs(\epsilon;t) = \sum_\nu\frac{\ket{\nust}\bra{\nust}}{\epsilon -\epsnut+i\eta}
  \label{eq:Gs_defn}
\end{equation}
with $\eta$ a positive infinitesimal.
By introducing the basis set $\ket{\bs}$ into (\ref{eq:elec_spec_dyn_2}) we find
\begin{equation}
  \ns(\epsilon,t) = -\frac{1}{\pi} \textrm{Im}\left\{\sum_{b,b'} n_{bb'\sigma}(t) \Gbbps(\epsilon;t)\right\}.
  \label{eq:elec_spec_dyn_2b}
\end{equation}
where $\Gbbps = \bra{\bs}\Gs\ket{\bps}$ and $\nbbps =
\bra{\bs}\hat{n}_{1\sigma}(t)\ket{\bps}$. $\nas(t)\equiv
n_{aa\sigma}(t)$ is the dynamically evolving adsorbate occupation
which appears in (\ref{eq:many_e_ham}) and (\ref{eq:ebar_defn}).
In our previous work\cite{mizielinski05} we obtained expressions
for $\nas(t)$, but here we also need the occupation functions
$\naks(t)$, $\nkas(t)$ and $\nkkps(t)$, as well as the full set of
instantaneous Green's functions.

The Green's functions can be found from the Dyson equation
\begin{equation}
  G(\epsilon;t) = G^0(\epsilon;t) + G^0(\epsilon;t)V(t)G(\epsilon;t),
  \label{eq:dyson}
\end{equation}
where $G^0$ is the unperturbed Green's function, and $V$ is the interaction potential. For our system we have
\begin{subequations}
  \begin{eqnarray}
    \Gaa(\epsilon;t) &=& \GaaI(\epsilon;t) + \GaaI(\epsilon;t)\sum_k \vak(t) \Gka(\epsilon;t), \\
    \Gak(\epsilon;t) &=& \GaaI(\epsilon;t)\sum_{k'} \vakp(t)\Gkpk(\epsilon;t), \\
    \Gka(\epsilon;t) &=& \GkkI(\epsilon;t) \vak^*(t) \Gaa(\epsilon;t), \\
    \Gkkp(\epsilon;t)&=& \GkkI(\epsilon;t)\delta_{k,k'} + \GkkI(\epsilon;t) \vak^*(t)\Gakp(\epsilon;t),
  \end{eqnarray}
  \label{eq:Greens_1}
\end{subequations}
where
\begin{subequations}
  \begin{eqnarray}
    \GaaI(\epsilon;t) &=& \frac{1}{\epsilon-\ebar(t)+i\eta},\\
    \GkkI(\epsilon;t) &=& \frac{1}{\epsilon-\eks+i\eta}.
  \end{eqnarray}
  \label{eq:Greens_0}
\end{subequations}
In order to find solutions to these equations we make two standard assumptions\cite{anderson61,mizielinski05}.
First we assume that the interaction potential $\vak$ can be separated into a complex constant and a real,
state-independent, time-varying function.
Second we assume that the resonance width $\Gamma$, defined as
\begin{equation}
  \Gamma(t) = 2\pi\sum_k \vert\vak(t)\vert^2\delta(\epsilon-\eks)
\label{eq:gamma_defn}
\end{equation}
is independent of energy $\epsilon$.
This is often referred to as the wide-band limit.
Using these approximations (\ref{eq:Greens_1}) becomes
\begin{subequations}
  \begin{eqnarray}
    \Gaa(\epsilon;t) &=& \frac{1}{\epsilon-\etilde(t)},
    \label{eq:Gaa}\\
    \Gak(\epsilon;t) &=& \frac{1}{(\epsilon-\etilde(t))}.\frac{\vak(t)}{(\epsilon-\eks+i\eta)},
    \label{eq:Gak}\\
    \Gka(\epsilon;t) &=& \frac{1}{(\epsilon-\etilde(t))}.\frac{\vak^\ast(t)}{(\epsilon-\eks+i\eta)},
    \label{eq:Gka}\\
    \Gkkp(\epsilon;t) &=& \frac{\delta_{kk'}}{\epsilon-\eks+i\eta} + \frac{\vak^{\ast}(t)}{(\epsilon-\eks+i\eta)}. \frac{1}{(\epsilon-\etilde(t))}. \frac{\vakp(t)}{(\epsilon-\epsilon_{k'\sigma}+i\eta)},
  \label{eq:Gkkp}
  \end{eqnarray}
  \label{eq:Greens_final}
\end{subequations}
where
\begin{equation}
  \etilde(t) = \ebar(t) -\frac{i}{2}\Gamma(t).
  \label{eq:etilde_defn}
\end{equation}
These Green's functions are very similar to those obtained by Anderson\cite{anderson61}, with the
exception of the use of the instantaneous energy level $\ebar(t)$ rather than the adiabatic level.

Substituting (\ref{eq:Greens_final}) into (\ref{eq:elec_spec_dyn_2b}),
and noting that $\naks = \nkas^*$,  allows us to write
\begin{eqnarray}
  \ns(\epsilon,t)\!\! &=&\!\! \nas(t)\rhoinst(\epsilon,t) + \sum_k \nkks(t) \delta(\epsilon-\eks) \nonumber\\
  &&\!\! -\frac{1}{\pi} \Im\left\{ \frac{1}{\epsilon-\etilde(t)} \int \frac{d\epsilon'}{\epsilon-\epsilon'+i\eta}.
  2\Re\left[\sum_k\naks(t)\vak(t)\delta(\epsilon'-\eks)\right]\right\} \nonumber \\
  &&\!\! - \frac{1}{\pi} \Im \Bigg\{\frac{1}{\epsilon-\etilde(t)} \int \frac{d\epsilon'}{\epsilon-\epsilon'+i\eta}
  \int \frac{d\epsilon''}{\epsilon-\epsilon''+i\eta} \nonumber  \\
  && \hspace{3cm} \times \sum_{k,k'} \nkkps(t)\vak^*(t)\vakp(t)\delta(\epsilon'-\eks)\delta(\epsilon''-\eksp) \Bigg\},
  \label{eq:elec_spec_dyn_4}
\end{eqnarray}
where we have introduced energy integrals to make the recovery of the width $\Gamma(t)$ easier.
$\rhoinst(\epsilon,t)$ is defined as
\begin{eqnarray}
  \rhoinst(\epsilon,t) &=& \frac{\Gamma(t)}{2\pi[(\epsilon-\ebar(t))^2 + \Gamma(t)^2/4]},
  \label{eq:rhoinst_defn}
\end{eqnarray}
which is very similar to the familiar adsorbate projected density of states, with the exception that once again
the energy $\ebar(t)$ is the instantaneous rather than the adiabatic level.

In order to complete our expression for $\ns$ we require $\nas$,
$\naks$ and $\nkkps$. Previously, we calculated $\nas$ by solving the equation of
motion for the creation and annihilation operators in the
Heisenberg picture\cite{mizielinski05}. This solution is given by
(we assume $\hbar = 1$):
\begin{eqnarray}
  \ca(t) &=& -i \int^t_{t_0} dt_1 \exp\left[-i\int^t_{t_1} \etilde(t')dt'\right]
  \sum_k \vak(t_1) \exp\left[-i\eks(t_1-t_0)\right]\ck(t_0) \nonumber \\
         & & + \exp\left[-i\int^t_{t_0} \etilde(t')dt'\right]\ca(t_0),
  \label{eq:ca_defn} \\
  \ck(t) &=& -i \int^t_{t_0} dt_1 \vak^*(t_1) \exp\left[-i\eks(t-t_1)\right]\ca(t_1)
  + \exp\left[-i\eks(t-t_0)\right]\ck(t_0),
  \label{eq:ck_defn}
\end{eqnarray}
from which we found
\begin{equation}
  \nas(t) \equiv \langle\!\langle \cad(t)\ca(t) \rangle\!\rangle
  = \nas(t_0)\exp\!\left[-\int^t_{t_0} \Gamma(t')dt'\right]
  + \int d\epsilon f(\epsilon)\vert\ps(\epsilon,t)\vert^2,
  \label{eq:nas_defn}
\end{equation}
where $\nas(t_0)=f(\ebar(t_0))$ is the initial adsorbate
occupation and $\ps$ is defined as
\begin{equation}
  \ps(\epsilon,t) = \int^t_{t_0} dt_1 \gpi{t_1} \exp\left[-i\int^t_{t_1}(\etilde(t')-\epsilon)dt'\right].
  \label{eq:ps_defn}
\end{equation}
As shown in Appendix \ref{app:occ_funcs} the remaining occupation functions can be evaluated in a similar manner 
using (\ref{eq:ca_defn}) and (\ref{eq:ck_defn}). Alternatively, the result for the occupation functions can be derived 
from (\ref{eq:onematrixresult}) and solving for the time evolution
of the $\ket{\mus(t)}$ states.

The distribution function $\ns$ now follows from (\ref{eq:elec_spec_dyn_4}) and the occupation functions in Appendix \ref{app:occ_funcs}.
The algebra is lengthy, but straightforward, with the definition of $\Gamma(t)$ (\ref{eq:gamma_defn}) used where necessary.
We find
\begin{eqnarray}
  \ns(\epsilon,t) \!\!&=& \!\!
  \int d\epsilon' f(\epsilon') \left| \qs(\epsilon,\epsilon',t)+\Pinst(\epsilon,t)\ps(\epsilon',t)\right|^2
  - 2f(\epsilon) \Re\lbrace\qs(\epsilon,\epsilon,t)\rbrace \nonumber \\
  & &\!\! + \nas(t_0) \left| \rs(\epsilon,t)+\Pinst(\epsilon,t) \exp\left[-\frac{1}{2}\int^t_{t_0} \Gamma(t')dt'\right]
  \right|^2 + \sum_k f(\eks)\delta(\epsilon-\eks) \nonumber \\
  & & \!\!+\frac{2}{\pi} \int d\epsilon' f(\epsilon') \Im\lbrace\ps(\epsilon',t)\rbrace
  \Re\left\{\frac{\Pinst(\epsilon,t)}{\epsilon-\epsilon'+i\eta}\right\}\nonumber \\
  &&-2\gpi{t} \int d\epsilon' f(\epsilon') \Im\left\{\frac{\qs^*(\epsilon,\epsilon',t)
    \Pinst(\epsilon,t)}{\epsilon-\epsilon'+i\eta}\right\}\nonumber \\
  & &\!\! - \frac{1}{\pi} \gpi{t}\int d\epsilon' f(\epsilon') \Re\left\{ \frac{\Pinst(\epsilon,t)}{
    (\epsilon-\epsilon'+i\eta)^2}\right\},
  \label{eq:ns_final}
\end{eqnarray}
where $\qs$, $\rs$ and $\Pinst$ are defined as
\begin{eqnarray}
  \qs(\epsilon,\epsilon',t) &=& \int^t_{t_0} dt_1 \gpi{t_1} \ps(\epsilon',t_1)\exp[i(\epsilon-\epsilon')(t_1-t)],
  \label{eq:qs_defn}\\
  \rs(\epsilon,t) &=& \exp\left[-\frac{1}{2}\int^t_{t_0} \Gamma(t')dt'\right] \int^t_{t_0} dt_1 \gpi{t_1}
  \exp\left[i\int^t_{t_1}(\etilde(t')-\epsilon)dt'\right],
  \label{eq:rs_defn}\\
  \Pinst(\epsilon,t) &=& \gpi{t}\frac{i}{\epsilon-\etilde(t)},
  \label{eq:Pinst_defn}
\end{eqnarray}
such that, from (\ref{eq:rhoinst_defn}),
\begin{eqnarray}
  \rhoinst(\epsilon,t) &=& \vert\Pinst(\epsilon,t)\vert^2.
\end{eqnarray}
Each of these quantities, along with $\nas$ and $\ps$ defined earlier, can be calculated numerically, as will be described
in section \ref{sec:computation}.
Given the variation of $\epsilon_a$ and $\Gamma$, and the value of $U$, the evolution of $\ns$ can therefore be computed.

There are two tests which we can perform on the distribution $\ns$ to confirm that it is correct.
First we can check that the number of electrons is conserved throughout the evolution of $\ns$ by integrating over all energies
$\epsilon$; in appendix \ref{app:charge_cons} we demonstrate that (\ref{eq:ns_final}) obeys charge conservation.
The second test is to calculate the total energy of the system by taking the first moment of the distribution function
(\ref{eq:elec_spec_dyn_energy_1}) and comparing this to the rate of change of energy derived previously\cite{mizielinski05}.
The verification that these approaches give the same result is long winded and will be presented in a future
publication\cite{thesis}.

\section{The excitation spectrum}
\label{sec:ad_spec}
\newcommand{\Gbbs}{G^{\sigma}_{bb}}
\newcommand{\epsnutz}{\epsilon^\nu_{\sigma t_0}}
\newcommand{\Depsnut}{\Delta\epsilon^\nu_{\sigma t}}
\newcommand{\Nk}{N^{metal}_\sigma}
\newcommand{\nsx}{n_\sigma^{(ex)}}
\newcommand{\nst}{n_{\sigma t}}

In order to analyse the excitations of our system it is important to recall the definition of
the distribution function $\ns(\epsilon,t)$; it is the time-evolving distribution of occupied electronic states.
However, the quantity we require is the spectrum of {\it excitations}, for which we need to
subtract an underlying distribution in which there are no electronic excitations. We write this
instantaneous distribution as $\nst(\epsilon)$; it is given by (\ref{eq:elec_spec_dyn}) with $\ket{\mus(t)}=\ket{\nust}$, which gives
\begin{equation}
  \nst(\epsilon) = \sum_\nu f(\epsnutz)\delta(\epsilon-\epsnut).
  \label{eq:nst_defn}
\end{equation}
The difference between the times that appear in this expression is significant. $f(\epsnutz)$ is the initial occupation
of the state that is connected to $\ket{\nust}$; this occupation does not change with $t$. In a finite system, however,
the eigenvalues can vary. In a system of $N$ electrons, the change in the eigenvalue $\Depsnut=\epsnut-\epsnutz$
will be of order $1/N$. We expand the Fermi function in \eqref{eq:nst_defn} to first order in $\Depsnut$, giving
\begin{eqnarray}
  \nst(\epsilon) &=& \sum_\nu \left[f(\epsnut) - \Depsnut \frac{df}{d\epsilon}(\epsnut)\right]\delta(\epsilon-\epsnut) \nonumber\\
    &=& f(\epsilon) \sum_\nu \delta(\epsilon-\epsnut) - \frac{df}{d\epsilon}(\epsilon)\sum_\nu \Depsnut \delta(\epsilon-\epsnut).
  \label{eq:nst_fermi_approx}
\end{eqnarray}
When integrated to give the total number of electrons in the sytem (as in Appendix B), the first term in this expression gives
a quantity of order $N$ electrons, while the second term integrates to a charge of order 1 electron (the sum over $\nu$
contains $N$ terms each of order $1/N$). In order to have an underlying distribution that conserves charge, it follows that
this second term cannot be neglected in the limit as $N\rightarrow \infty$. Higher order terms in the expansion of the
Fermi function will yield of order $1/N$ electrons or less, and in the $N\rightarrow \infty$ limit these terms can be ignored.

The first term in \eqref{eq:nst_fermi_approx} can be dealt with using the Green's functions derived in the previous section.
By introducing the basis states $\ket{\bs}$ the first term in \eqref{eq:nst_fermi_approx} becomes
\begin{eqnarray}
  f(\epsilon)\sum_{b,\nu} \vert\braket{\bs}{\nust}\vert^2\delta(\epsilon-\epsnut) =
  -\frac{f(\epsilon)}{\pi} \sum_b \Im\left\{\Gbbs(\epsilon;t)\right\}.
\label{eq:nst_2}
\end{eqnarray}
When combined with \eqref{eq:Gaa} and \eqref{eq:Gkkp}, and using the wide-band approximation, this yields
\begin{equation}
  f(\epsilon)\rhoinst(\epsilon,t) + \sum_k f(\eks)\delta(\epsilon-\eks),
  \label{eq:nst_3}
\end{equation}
and therefore
\begin{equation}
  \nst(\epsilon) = f(\epsilon)\rhoinst(\epsilon,t) + \sum_k f(\eks) \delta(\epsilon-\eks) - \frac{df}{d\epsilon}
  \sum_\nu \Depsnut \delta(\epsilon-\epsnut).
\label{eq:nst_4}
\end{equation}
A more convenient form for the third term in \eqref{eq:nst_4} can be determined by consideration of charge conservation.
As in Appendix B, the integral of the instantaneous distribution function over energy:
\begin{equation}
  \int d\epsilon \phantom{l}\nst(\epsilon)= \int d\epsilon f(\epsilon) \rhoinst(\epsilon,t) + \sum_k f(\eks) -
  \int d\epsilon \frac{df}{d\epsilon}\sum_\nu \Depsnut \delta(\epsilon-\epsnut),
  \label{eq:nst_chg}
\end{equation}
should give the total number of electrons of spin $\sigma$ in the system. The second term here gives the initial number of metal 
electrons, and so to conserve the number of electrons in $\nst(\epsilon)$ we must have
\begin{equation}
-\int d\epsilon \frac{df}{d\epsilon} \sum_\nu \Depsnut \delta(\epsilon-\epsnut) =
\nas(t_0) - \int d\epsilon f(\epsilon) \rhoinst(\epsilon,t).
\label{eq:nst_chg_2}
\end{equation}
We now assume that the sum over the states $\ket{\nust}$ in \eqref{eq:nst_chg_2} is independent of energy over a
range of several $k_BT$ either side of the Fermi level.
This assumption is consistent with the wide-band approximation (see \eqref{eq:gamma_defn}) used throughout this work.
Equation \eqref{eq:nst_chg_2} then gives
\begin{equation}
    \sum_\nu \Depsnut \delta(\epsilon-\epsnut) = \nas(t_0) - \int d\epsilon f(\epsilon) \rhoinst(\epsilon,t)
\label{eq:nst_chg_3}
\end{equation}
and consequently the instantaneous distribution function becomes
\begin{equation}
\nst(\epsilon)=f(\epsilon)\rhoinst(\epsilon,t) + \sum_k f(\eks) \delta(\epsilon-\eks) - \frac{df}{d\epsilon}
\left[\nas(t_0) - \int d\epsilon' f(\epsilon') \rhoinst(\epsilon',t)\right].
\label{eq:nst_final}
\end{equation}
The difference between $\ns(\epsilon,t)$, \eqref{eq:ns_final}, and $\nst(\epsilon)$, \eqref{eq:nst_final}, is
the required spectrum of excitations $\nsx(\epsilon,t)$.

\section{Computational methods}
\label{sec:computation}
\newcommand{\PVa}{\textrm{PV}_\alpha}
\newcommand{\nsD}{n_{\sigma D}}
\newcommand{\nsxD}{n_{\sigma D}^{(ex)}}

In this section we outline the methods we have used to compute the distribution functions for the time-evolving, (\ref{eq:ns_final}),
and instantaneous, (\ref{eq:nst_final}), systems.
To compute $\ns(\epsilon,t)$ we require three quantities; $\ps$, $\qs$ and $\rs$.
As in reference \onlinecite{mizielinski05} $\ps$, (\ref{eq:ps_defn}), is obtained from
\begin{equation}
  \frac{d\ps}{dt}(\epsilon',t) = -i(\etilde(t) - \epsilon')\ps(\epsilon',t) + \gpi{t},
  \label{eq:dp_dt}
\end{equation}
with the initial condition $\ps(\epsilon',t_0)=0$. This equation is integrated on a finite grid of $\epsilon'$ points using the fourth order
Runge-Kutta method.
The evolution of the system is driven by variation of $\epsilon_a$ and $\Gamma$ with time, with the initial condition that $\Gamma(t_0)=0$.
The energy $\ebar$ (\ref{eq:ebar_defn}), and hence $\etilde$ (\ref{eq:etilde_defn}), is calculated using $\nas$ which is obtained
from (\ref{eq:nas_defn}).
A sufficiently fine $\epsilon'$ grid is required to ensure the accuracy of the energy integral in (\ref{eq:nas_defn}).
$\qs$ is obtained from $\ps$ by using
\begin{equation}
  \qs(\epsilon,\epsilon',t)=\exp[-i(\epsilon-\epsilon')(t-t_0)] \int^t_{t_0} dt_1 \gpi{t_1} \ps(\epsilon',t_1)\exp[i(\epsilon-\epsilon')(t_1-t_0)].
  \label{eq:qs_defn_comp}
\end{equation}
The $\epsilon$ variable here is represented by a set of energy values at which the distribution function $\ns(\epsilon,t)$ is required.
The time integral in (\ref{eq:qs_defn_comp}) is performed numerically using Simpson's method
for each $(\epsilon',\epsilon)$ grid point combination.
The removal of any $t$ dependence from the integrand in (\ref{eq:qs_defn_comp}) makes it possible to evaluate $\qs$ at regular points in
the Simpson's integration, rather than at a predefined time $t$.
We use this property to explore the evolution of the distribution function $\ns$, as $\ebar$ and $\Gamma$ vary.
$\rs$, (\ref{eq:rs_defn}), is found  in a similar manner to $\ps$ from
\begin{equation}
   \frac{d\rs}{dt} (\epsilon,t) = i(\ebar(t)-\epsilon)\rs(\epsilon,t) + \gpi{t}\exp\left[-\frac{1}{2}\int^t_{t_0}\Gamma(t')dt'\right],
  \label{eq:dr_dt}
\end{equation}
with the initial condition $\rs(\epsilon,t_0)=0$.
The integration of $\rs$ only needs to be performed for the energies $\epsilon$.

Using (\ref{eq:dp_dt}), (\ref{eq:qs_defn_comp}), (\ref{eq:dr_dt}) and the definition of $\nas$ (\ref{eq:nas_defn}),
the first three terms in $\ns(\epsilon,t)$ can be calculated. The fourth term in $\ns$ cancels with the second term in the
instantaneous distribution function $\nst$, (\ref{eq:nst_final}), and can therefore be ignored.
The final three terms in (\ref{eq:ns_final}) require further work due to the singularity in their
integrands in the $\eta\rightarrow0^+$ limit.

The $\epsilon'$ integral in each of the final three terms is separated into three sections; a window of width
$2\alpha$ around $\epsilon'=\epsilon$, a section below $\epsilon'=\epsilon-\alpha$ and a section above $\epsilon'=\epsilon+\alpha$.
The sum of the outer sections is similar to a principal value integral and we therefore use the notation $\PVa$ to denote this.
We then Taylor expand the integrand in the central window and perform this section of the integral analytically.
We will demonstrate this `analytic window' approximation using the fifth term in (\ref{eq:ns_final}),
$\ns^{(5)}$, and then state the results for the sixth and seventh terms, $\ns^{(6)}$ and $\ns^{(7)}$.

The $\epsilon'$ integral in $\ns^{(5)}$ becomes
\begin{eqnarray}
  \ns^{(5)}(\epsilon, t) &=& \frac{2}{\pi}\Re\left\{\Pinst(\epsilon,t)\right\}
  \PVa \int d\epsilon' \frac{f(\epsilon')}{\epsilon-\epsilon'} \Im\left\{\ps(\epsilon',t)\right\}\nonumber\\
  &&+\frac{2}{\pi} \int^{\epsilon+\alpha}_{\epsilon-\alpha} d\epsilon' f(\epsilon')
  \Im\left\{\ps(\epsilon',t)\right\}\Re\left\{\frac{\Pinst(\epsilon,t)}{\epsilon-\epsilon'+i\eta}\right\},
  \label{eq:ns5_window_1}
\end{eqnarray}
where we have dropped the $\eta$ from the first term as the $\PVa$ integral does not cover the region in which it is important.
We now expand the product $f(\epsilon')\ps(\epsilon',t)$ as a Taylor series about $\epsilon'=\epsilon$ to first order, yielding
\begin{eqnarray}
  \ns^{(5)}(\epsilon,t)&=&\frac{2}{\pi}\Re\left\{\Pinst(\epsilon,t)\right\}
  \PVa \int d\epsilon \frac{f(\epsilon')}{\epsilon-\epsilon'}\Im\left\{\ps(\epsilon',t)\right\} \nonumber \\
  & & + \frac{2}{\pi}\Re\left\{\Pinst(\epsilon,t) \int^{\epsilon+\alpha}_{\epsilon-\alpha}
  \frac{d\epsilon'}{\epsilon-\epsilon'+i\eta}\right\} f(\epsilon)\Im\left\{\ps(\epsilon,t)\right\} \nonumber \\
  & & + \frac{2}{\pi}\Re\left\{\Pinst(\epsilon,t)\int^{\epsilon+\alpha}_{\epsilon-\alpha} d\epsilon'
  \frac{\epsilon-\epsilon'}{\epsilon-\epsilon' + i\eta}\right\}
  \frac{d}{d\epsilon}\Big[f(\epsilon)\Im\left\{\ps(\epsilon,t)\right\}\Big] .
  \label{eq:ns5_window_2}
\end{eqnarray}
The integrals in the second and third terms in (\ref{eq:ns5_window_2}) can be evaluated and taking the $\eta\rightarrow0^+$ limit we obtain
\begin{eqnarray}
  \ns^{(5)}(\epsilon,t) &=& 2 \gpi{t} \rhoinst(\epsilon,t) \textrm{PV}_\alpha \int d\epsilon' \frac{f(\epsilon')}{\epsilon-\epsilon'}
  \Im\left\{\ps(\epsilon',t)\right\} \nonumber \\
  & & + 2\sqrt{\frac{2\pi}{\Gamma(t)}} (\epsilon-\ebar(t)) \rhoinst(\epsilon,t) f(\epsilon)\Im\{\ps(\epsilon,t)\} \nonumber \\
  & & - 4\alpha \gpi{t} \rhoinst(\epsilon,t) \left(\frac{df}{d\epsilon}\,\Im\{\ps(\epsilon,t)\}
  + f(\epsilon)\Im\left\{\frac{d\ps}{d\epsilon}(\epsilon,t)\right\}\right),
  \label{eq:ns5_window}
\end{eqnarray}
where we have used the definitions of $\rhoinst$, (\ref{eq:rhoinst_defn}), and $\Pinst$, (\ref{eq:Pinst_defn}).
The energy derivative of $\ps$ is obtained using a finite centred difference method.
We note here that this expansion includes all first order terms in $\alpha$, and will give the exact result in the $\alpha\rightarrow0$ limit.

By using the same method for $\ns^{(6)}$ we find
\begin{eqnarray}
  &&\hspace{-1cm}\ns^{(6)}(\epsilon,t) =\nonumber \\
  &&-2\gpi{t}\PVa\int d\epsilon' \frac{f(\epsilon')}{\epsilon-\epsilon'} \Im\left\{\qs^*(\epsilon,\epsilon',t) \Pinst(\epsilon,t)\right\} \nonumber \\
  && + \left(2\pi - 4\alpha(t-t_0)\right)\gpi{t}f(\epsilon)\Re\left\{\qs^*(\epsilon,\epsilon,t)\Pinst(\epsilon,t)\right\} \nonumber \\
  && + 4\alpha \gpi{t} \frac{df}{d\epsilon}\,\Im\left\{\qs^*(\epsilon,\epsilon,t)\Pinst(\epsilon,t)\right\} \nonumber \\
  && + 4\alpha \gpi{t} f(\epsilon)\Im\left\{\Pinst(\epsilon,t) \int_{t_0}^t dt_1\gpi{t_1}\left(\frac{d\ps^*}{d\epsilon}(\epsilon,t_1)
  + i(t_1-t_0) \ps^*(\epsilon,t)\right)\right\}.
  \label{eq:ns6_window}
\end{eqnarray}
The final term in (\ref{eq:ns_final}), $\ns^{(7)}$, requires a little more work before applying our approximation due to the
$1/(\epsilon-\epsilon')^2$ dependence of the integrand. Integrating $\ns^{(7)}$ by parts yields
\begin{eqnarray}
  \ns^{(7)}(\epsilon,t) &=&
  -\frac{1}{\pi}\gpi{t}\Re\left\{\Pinst(\epsilon,t) \left(\left[\frac{-f(\epsilon')}{\epsilon-\epsilon'+i\eta}\right]^{\epsilon'_+}_{\epsilon'_-}
  + \int d\epsilon' \frac{df}{d\epsilon'} \frac{1}{\epsilon-\epsilon'+i\eta}\right)\right\}\nonumber \\
  &=& -\frac{\Gamma(t)}{2\pi}.\frac{\rhoinst(\epsilon,t)}{(\epsilon-\epsilon'_-)}
  -\frac{1}{\pi}\gpi{t}\int d\epsilon' \frac{df}{d\epsilon'} \Re\left\{\frac{\Pinst(\epsilon,t)}{\epsilon-\epsilon'+i\eta}\right\},
  \label{eq:ns7_window_1}
\end{eqnarray}
where $\epsilon'_+$ and $\epsilon'_-$ are the upper and lower limits of the $\epsilon'$ range over
which we are integrating, and we have assumed
that $f(\epsilon'_+)=0$ and $f(\epsilon'_-)=1$.
The remaining integral in (\ref{eq:ns7_window_1}) can be dealt with in the same manner as $\ns^{(5)}$ and $\ns^{(6)}$, giving
\begin{eqnarray}
  \ns^{(7)}(\epsilon,t) &=& -\frac{\Gamma(t)}{2\pi}.\frac{\rhoinst(\epsilon,t)}{(\epsilon-\epsilon'_-)}
  - \frac{\Gamma(t)}{2\pi}\rhoinst(\epsilon,t)
  \PVa \int d\epsilon'\frac{df}{d\epsilon'} \frac{1}{\epsilon-\epsilon'} \nonumber \\
  & & - (\epsilon-\ebar(t))\frac{df}{d\epsilon} \rhoinst(\epsilon,t)
  + \frac{\alpha\Gamma(t)}{\pi}\frac{d^2f}{d\epsilon^2}\rhoinst(\epsilon,t).
  \label{eq:ns7_window}
\end{eqnarray}
This result, like (\ref{eq:ns5_window}) and (\ref{eq:ns6_window}), includes all first order terms in the window half-width $\alpha$.

The instantaneous distribution function is more straightforward to calculate than $\ns$.
The first term in $\nst$, \eqref{eq:nst_final}, can be calculated directly using the definition of $\rhoinst$, \eqref{eq:rhoinst_defn}.
The second term in \eqref{eq:nst_final} cancels with an identical term in $\ns$, \eqref{eq:ns_final}.
The final term in \eqref{eq:nst_final} cannot be used directly due to the truncation of the $\epsilon'$ range over which we
perform numerical integrals. We require the integral of $\nst$ over all energies $\epsilon$ to conserve the number of
electrons and consequently we modify the final term in \eqref{eq:nst_final} to
\begin{eqnarray}
  &&\hspace{-1.5cm}-\frac{df}{d\epsilon} \left[\nas(t_0) - \int^\infty_{-\infty} d\epsilon f(\epsilon) \rhoinst(\epsilon,t)\right]\nonumber \\
  &=&\!\!\! -\frac{df}{d\epsilon}\left[\nas(t_0) -\int^{\epsilon'_+}_{\epsilon'_-}d\epsilon' f(\epsilon')\rhoinst(\epsilon') -\int^{\epsilon'_-}_{-\infty} d\epsilon \rhoinst(\epsilon,t)\right] \nonumber \\
  &=&\!\!\!-\frac{df}{d\epsilon}\Bigg[\nas(t_0) -\int^{\epsilon'_+}_{\epsilon'_-} d\epsilon' f(\epsilon')\rhoinst(\epsilon',t) 
-\left(\frac{1}{2}-\frac{1}{\pi}\tan^{-1}\left\{2\frac{\ebar(t)-\epsilon'_-}{\Gamma(t)}\right\}\right)\Bigg]\!,
  \label{eq:num_nsad_fix}
\end{eqnarray}
where we have assumed $f(\epsilon'_+)=0$ as before and have performed the integral over $\epsilon'$ up to $\epsilon'_-$ analytically.

It is convenient when performing numerical calculations to work with de-dimensionalised parameters.
As in our previous work\cite{mizielinski05} we scale all parameters by a width $\Gamma_0$; so that, for example,
$t_D=\Gamma_0 t$, $\Gamma_D=\Gamma/\Gamma_0$, $\nsD=\Gamma_0 \ns$, $\epsilon_D = \epsilon/\Gamma_0$, $U_D=U/\Gamma_0$
 and $k_BT_D=k_BT/\Gamma_0$ where we have used subscript $D$ to denote the dimensionless quantities.
Typically in practice $\Gamma_0$ will have a value between 1 and 3 eV.

We have carried out extensive tests of the numerical stability of our calculations.
We find that for run times of up to 200 dimensionless time units, well converged results are obtained for 160,001 $\epsilon'_D$ points in
the range $-$20 to 20, where the Fermi level is fixed at $\epsilon'_D=0$.
The analytic window half-width $\alpha$ is set to be the same as the $\epsilon'_D$ grid spacing and the Runge-Kutta integration step is 0.01
dimensionless time units.
The use of a non-zero temperature in our model aids the stability of $\nsD$, particularly in the region around the Fermi level.
In this work we have used a value of $k_BT_D=0.02$, which typically corresponds to a temperature of the order of a few hundred Kelvin.
With these parameters we have found that our computed distribution conserves charge over the full length of a
dynamical run to better than 10$^{-4}$ of an electron.

\section{Numerical results}
\label{sec:num_res}
\newcommand{\eaD}{\epsilon_{aD}}
We have chosen three model systems to demonstrate the non-adiabatic behaviour of the \NA{} system.
In each case $U_D=3$ and $\eaD$ is held fixed.
The model is driven by varying $\Gamma_D$ from 0 to 3 over 50 dimensionless time units using an error function with a peak gradient of
$d\Gamma_D/dt_D=0.3$ at $t_D=25$.
The three model systems we have chosen differ in their bare energy levels $\epsilon_{aD}$;
we use $\eaD=-$2.5, $-$1.5 and $-$0.5.
Each of these systems is driven through the spin transition by the $\Gamma_D$ variation.
The first system has the minority spin energy level closest to the Fermi level $\epsilon_F$, the second has the energy levels $\ebar$
equidistant from $\epsilon_F$ and the final system has the majority spin energy level closest to $\epsilon_F$.
These parameters were chosen to be broadly comparable with those extracted from DFT calculations of the H/Cu (111) system explored
previously\cite{mizielinski05}, with the exception that the rate of change of the parameter $\Gamma$ has been increased to exaggerate the
non-adiabatic behaviour.

\begin{figure}
  \begin{center}
    \begin{picture}(0,0)(-50,100)%
      \includegraphics{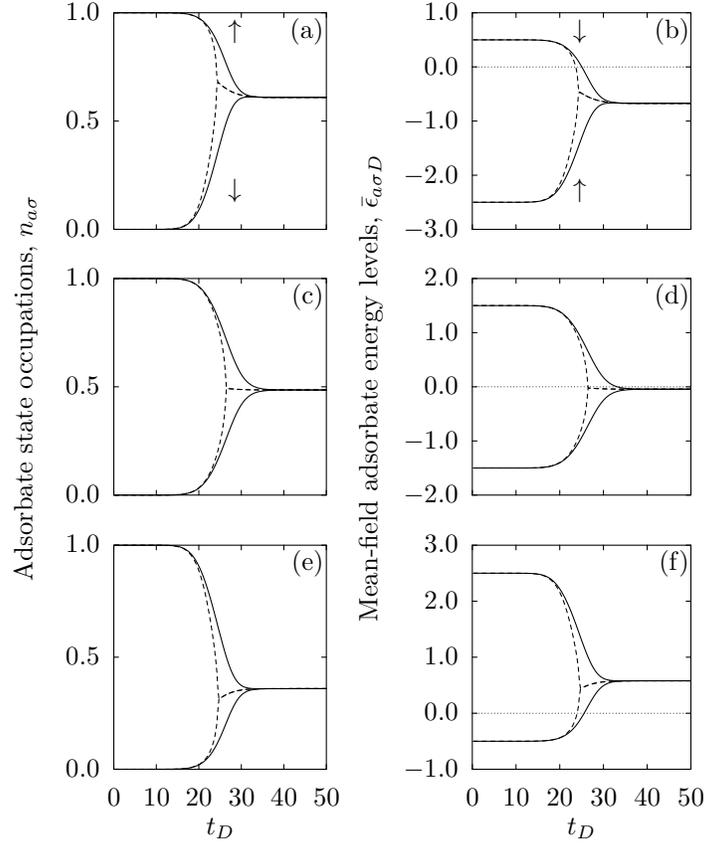}%
    \end{picture}%
    \begingroup
    \setlength{\unitlength}{0.0200bp}%
    \begin{picture}(18000,10800)(-2500,5000)%
      \put(1375,11281){\makebox(0,0)[r]{\strut{}0.0}}%
      \put(1375,13323){\makebox(0,0)[r]{\strut{}0.5}}%
      \put(1375,15365){\makebox(0,0)[r]{\strut{}1.0}}%
      \put(5600,15038){\makebox(0,0)[r]{\strut{}(a)}}%
      \put(3800,15038){\makebox(0,0)[l]{\strut{}$\uparrow$}}%
      \put(3800,12038){\makebox(0,0)[l]{\strut{}$\downarrow$}}%
      \put(0,8312){\rotatebox{90}{\makebox(0,0)[c]{\strut{}Adsorbate state occupations, $n_{a\sigma}$}}}%
      \put(1375,6270){\makebox(0,0)[r]{\strut{}0.0}}%
      \put(1375,8312){\makebox(0,0)[r]{\strut{}0.5}}%
      \put(1375,10354){\makebox(0,0)[r]{\strut{}1.0}}%
      \put(5600,10027){\makebox(0,0)[r]{\strut{}(c)}}%
      \put(1375,1100){\makebox(0,0)[r]{\strut{}0.0}}%
      \put(1375,3213){\makebox(0,0)[r]{\strut{}0.5}}%
      \put(1375,5325){\makebox(0,0)[r]{\strut{}1.0}}%
      \put(1650,550){\makebox(0,0){\strut{}0}}%
      \put(2451,550){\makebox(0,0){\strut{}10}}%
      \put(3252,550){\makebox(0,0){\strut{}20}}%
      \put(4053,550){\makebox(0,0){\strut{}30}}%
      \put(4854,550){\makebox(0,0){\strut{}40}}%
      \put(5655,550){\makebox(0,0){\strut{}50}}%
      \put(3652,0){\makebox(0,0)[c]{\strut{}$t_D$}}%
      \put(5600,4987){\makebox(0,0)[r]{\strut{}(e)}}%
      \put(8130,11281){\makebox(0,0)[r]{\strut{}$-$3.0}}%
      \put(8130,12302){\makebox(0,0)[r]{\strut{}$-$2.0}}%
      \put(8130,13323){\makebox(0,0)[r]{\strut{}$-$1.0}}%
      \put(8130,14344){\makebox(0,0)[r]{\strut{}0.0}}%
      \put(8130,15365){\makebox(0,0)[r]{\strut{}1.0}}%
      \put(12500,15038){\makebox(0,0)[r]{\strut{}(b)}}%
      \put(10300,15038){\makebox(0,0)[l]{\strut{}$\downarrow$}}%
      \put(10300,12038){\makebox(0,0)[l]{\strut{}$\uparrow$}}%
      \put(8130,6270){\makebox(0,0)[r]{\strut{}$-$2.0}}%
      \put(8130,7291){\makebox(0,0)[r]{\strut{}$-$1.0}}%
      \put(8130,8312){\makebox(0,0)[r]{\strut{}0.0}}%
      \put(8130,9333){\makebox(0,0)[r]{\strut{}1.0}}%
      \put(8130,10354){\makebox(0,0)[r]{\strut{}2.0}}%
      \put(6500,8312){\rotatebox{90}{\makebox(0,0)[c]{\strut{}Mean-field adsorbate energy levels, $\bar{\epsilon}_{a\sigma D}$}}}%
      \put(12500,10027){\makebox(0,0)[r]{\strut{}(d)}}%
      \put(8130,1100){\makebox(0,0)[r]{\strut{}$-$1.0}}%
      \put(8130,2156){\makebox(0,0)[r]{\strut{}0.0}}%
      \put(8130,3213){\makebox(0,0)[r]{\strut{}1.0}}%
      \put(8130,4269){\makebox(0,0)[r]{\strut{}2.0}}%
      \put(8130,5325){\makebox(0,0)[r]{\strut{}3.0}}%
      \put(8405,550){\makebox(0,0){\strut{}0}}%
      \put(9229,550){\makebox(0,0){\strut{}10}}%
      \put(10053,550){\makebox(0,0){\strut{}20}}%
      \put(10876,550){\makebox(0,0){\strut{}30}}%
      \put(11700,550){\makebox(0,0){\strut{}40}}%
      \put(12524,550){\makebox(0,0){\strut{}50}}%
      \put(10464,0){\makebox(0,0)[c]{\strut{}$t_D$}}%
      \put(12500,4987){\makebox(0,0)[r]{\strut{}(f)}}%
    \end{picture}%
    \endgroup
  \end{center}
  \vspace{3.5cm}
    \caption{Adsorbate state occupation and energy level variation for the adiabatic (dashed lines) and time-evolving (solid lines) models:
    (a), (c) and (e) occupations for $\epsilon_{aD}=-$2.5, $-$1.5 and $-$0.5 respectively,
    (b), (d) and (f) mean-field energy levels for $\epsilon_{aD}=-$2.5, $-$1.5 and $-$0.5 respectively.
    The dotted line in (b), (d) and (f) denotes the Fermi level. The arrows in (a) and (b) denote the spin-states,
    with $\uparrow$ indicating majority spin and $\downarrow$ minority spin.
    \label{fig:occ}}
\end{figure}

Figure \ref{fig:occ} shows the evolution the adsorbate state occupations and the mean-field energy levels, along
with the adiabatic equivalents calculated as in ref \onlinecite{mizielinski05}, for each of these model systems.
The $\eaD=-2.5$ calculations show the minority spin energy level crossing the Fermi level, gaining occupation and
converging to the majority level. The $\eaD=-1.5$ levels converge simultaneously on the final state, with small
deviations from the analytic result ($\nas=1/2$, $\ebar=\epsilon_F$)
due to the truncation of the $\epsilon'_D$ range in the numerical calculations.
The final model system, with $\eaD=-0.5$, involves the majority level crossing the Fermi level and approaching the
falling minority level, resulting in a low occupancy final state.
In each of the model systems the adiabatic occupations exhibit a sharp transition from a spin-polarised to a
spin-degenerate state at around $t_D=25$.
The dynamical occupations, however, overshoot this sharp spin transition and the net polarisation then
falls roughly exponentially to below 0.01 by $t_D=35$.

\begin{figure}
  \begin{center}
    \begin{picture}(0,0)(-40,100)%
      \includegraphics{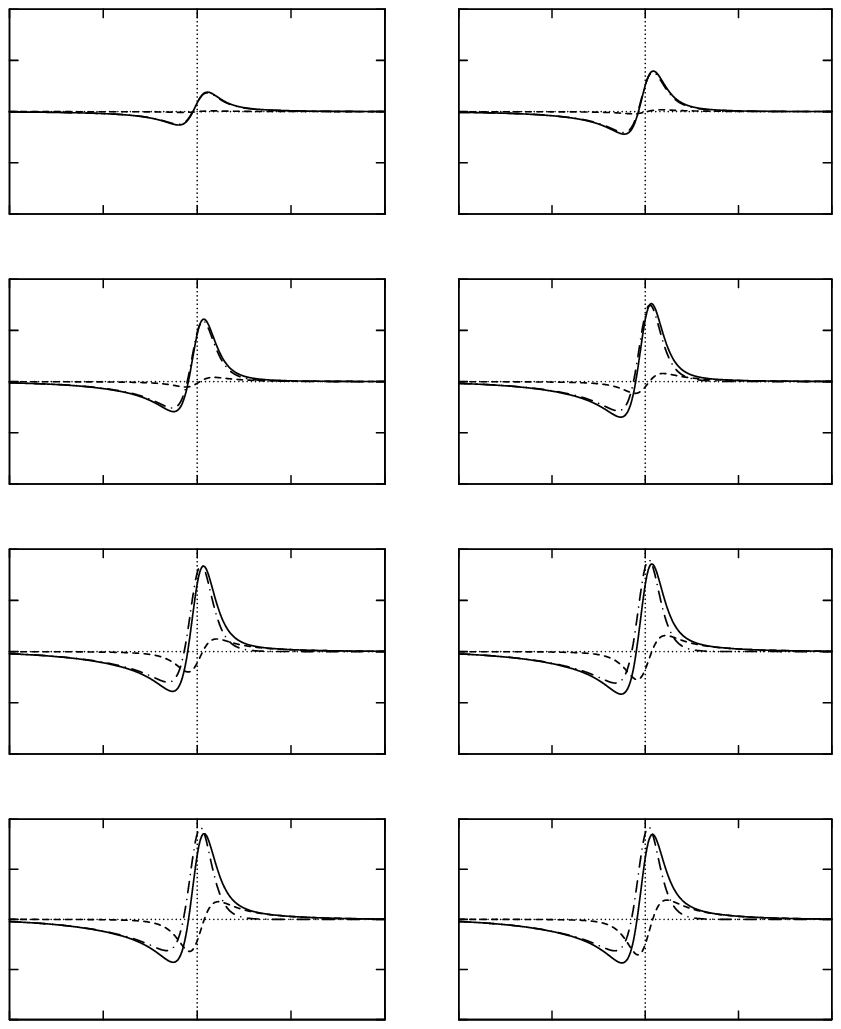}%
    \end{picture}%
    \begingroup
    \setlength{\unitlength}{0.0200bp}%
    \begin{picture}(18000,10800)(-2000,5000)%
      \put(1100,12700){\makebox(0,0)[r]{\strut{}$-$8}}%
      \put(1100,14175){\makebox(0,0)[r]{\strut{}0}}%
      \put(1100,15650){\makebox(0,0)[r]{\strut{}8}}%
      \put(6750,15300){\makebox(0,0)[r]{\strut{}(a)}}%
      \put(1100,8812){\makebox(0,0)[r]{\strut{}$-$8}}%
      \put(1100,10287){\makebox(0,0)[r]{\strut{}0}}%
      \put(1100,11762){\makebox(0,0)[r]{\strut{}8}}%
      \put(6750,11412){\makebox(0,0)[r]{\strut{}(c)}}%
      \put(1100,4924){\makebox(0,0)[r]{\strut{}$-$8}}%
      \put(1100,6399){\makebox(0,0)[r]{\strut{}0}}%
      \put(1100,7874){\makebox(0,0)[r]{\strut{}8}}%
      \put(6750,7524){\makebox(0,0)[r]{\strut{}(e)}}%
      \put(1100,1100){\makebox(0,0)[r]{\strut{}$-$8}}%
      \put(1100,2543){\makebox(0,0)[r]{\strut{}0}}%
      \put(1100,3986){\makebox(0,0)[r]{\strut{}8}}%
      \put(1375,550){\makebox(0,0){\strut{}$-$0.5}}%
      \put(4078,550){\makebox(0,0){\strut{}0.0}}%
      \put(6780,550){\makebox(0,0){\strut{}0.5}}%
      \put(0,8375){\rotatebox{90}{\makebox(0,0)[c]{\strut{}Excitation spectrum, $n_{\sigma_D}^{(ex)}$ }}}
      \put(4078,0){\makebox(0,0)[c]{\strut{}$\epsilon_D$}}%
      \put(6750,3642){\makebox(0,0)[r]{\strut{}(g)}}%
      \put(13188,15300){\makebox(0,0)[r]{\strut{}(b)}}%
      \put(13188,11412){\makebox(0,0)[r]{\strut{}(d)}}%
      \put(13188,7524){\makebox(0,0)[r]{\strut{}(f)}}%
      \put(7845,550){\makebox(0,0){\strut{}$-$0.5}}%
      \put(10530,550){\makebox(0,0){\strut{}0.0}}%
      \put(13215,550){\makebox(0,0){\strut{}0.5}}%
      \put(10530,0){\makebox(0,0)[c]{\strut{}$\epsilon_D$}}%
      \put(13188,3642){\makebox(0,0)[r]{\strut{}(h)}}%
    \end{picture}%
    \endgroup
  \end{center}
  \vspace{3.5cm}
  \caption{
    Excitation spectra snapshots for $\epsilon_{aD}=-2.5$ at times
    (a) $t_D=22.00$, (b) $t_D=23.25$, (c) $t_D=24.50$, (d) $t_D=25.75$, (e) $t_D=27.00$, (f) $t_D=28.25$, (g) $t_D=29.50$,
    and (h) after $\Gamma_D$ variation has finished ($t_D=50.00$). Solid lines denote the total electronic excitation spectrum,
    dashed the spin $\uparrow$ (majority) component and dot-dashed the spin $\downarrow$ (minority) component.
    \label{fig:ea-2_5}}
\end{figure}
\begin{figure}
  \begin{center}
    \begin{picture}(0,0)(-40,100)%
      \includegraphics{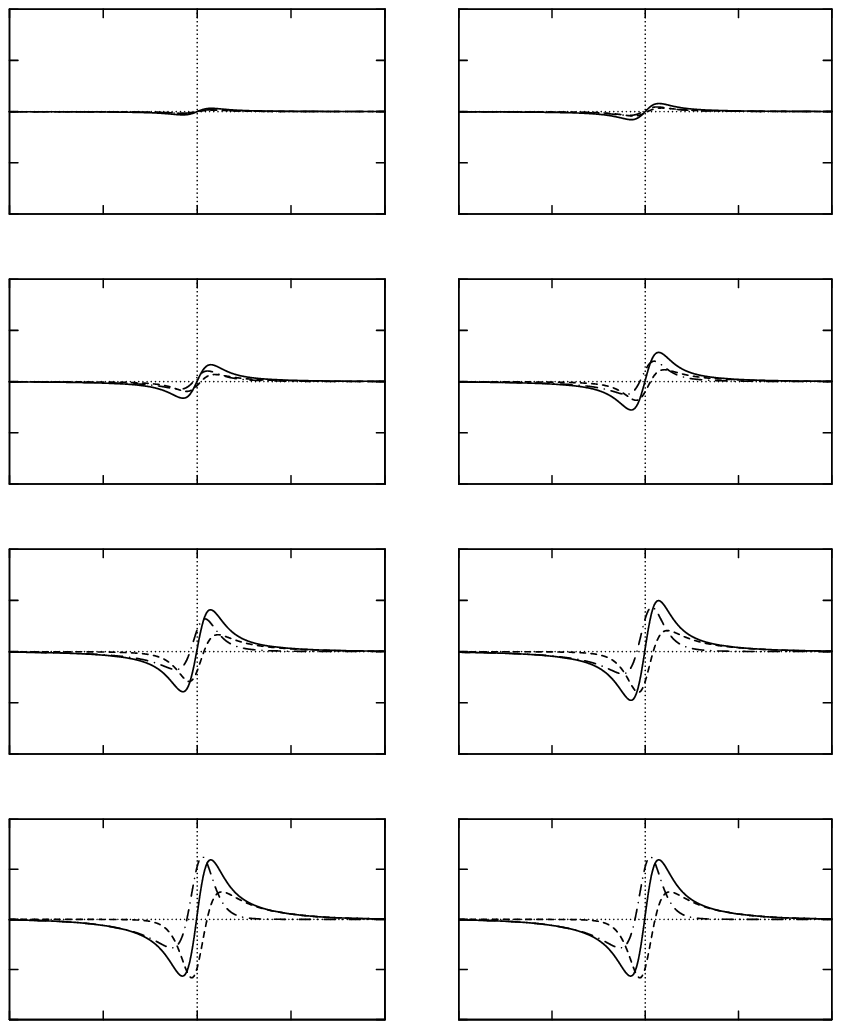}%
    \end{picture}%
    \begingroup
    \setlength{\unitlength}{0.0200bp}%
    \begin{picture}(18000,10800)(-2000,5000)%
      \put(1100,12700){\makebox(0,0)[r]{\strut{}$-$8}}%
      \put(1100,14175){\makebox(0,0)[r]{\strut{}0}}%
      \put(1100,15650){\makebox(0,0)[r]{\strut{}8}}%
      \put(6750,15300){\makebox(0,0)[r]{\strut{}(a)}}%
      \put(1100,8812){\makebox(0,0)[r]{\strut{}$-$8}}%
      \put(1100,10287){\makebox(0,0)[r]{\strut{}0}}%
      \put(1100,11762){\makebox(0,0)[r]{\strut{}8}}%
      \put(6750,11412){\makebox(0,0)[r]{\strut{}(c)}}%
      \put(1100,4924){\makebox(0,0)[r]{\strut{}$-$8}}%
      \put(1100,6399){\makebox(0,0)[r]{\strut{}0}}%
      \put(1100,7874){\makebox(0,0)[r]{\strut{}8}}%
      \put(6750,7524){\makebox(0,0)[r]{\strut{}(e)}}%
      \put(1100,1100){\makebox(0,0)[r]{\strut{}$-$8}}%
      \put(1100,2543){\makebox(0,0)[r]{\strut{}0}}%
      \put(1100,3986){\makebox(0,0)[r]{\strut{}8}}%
      \put(1375,550){\makebox(0,0){\strut{}$-$0.5}}%
      \put(4078,550){\makebox(0,0){\strut{}0.0}}%
      \put(6780,550){\makebox(0,0){\strut{}0.5}}%
      \put(0,8375){\rotatebox{90}{\makebox(0,0)[c]{\strut{}Excitation spectrum, $n_{\sigma D}^{(ex)}$}}}
      \put(4078,0){\makebox(0,0)[c]{\strut{}$\epsilon_D$}}%
      \put(6750,3642){\makebox(0,0)[r]{\strut{}(g)}}%
      \put(13188,15300){\makebox(0,0)[r]{\strut{}(b)}}%
      \put(13188,11412){\makebox(0,0)[r]{\strut{}(d)}}%
      \put(13188,7524){\makebox(0,0)[r]{\strut{}(f)}}%
      \put(7845,550){\makebox(0,0){\strut{}$-$0.5}}%
      \put(10530,550){\makebox(0,0){\strut{}0.0}}%
      \put(13215,550){\makebox(0,0){\strut{}0.5}}%
      \put(10530,0){\makebox(0,0)[c]{\strut{}$\epsilon_D$}}%
      \put(13188,3642){\makebox(0,0)[r]{\strut{}(h)}}%
    \end{picture}%
    \endgroup
  \end{center}
  \vspace{3.5cm}
  \caption{
    As for Fig. \ref{fig:ea-2_5}, but with $\eaD=-1.5$.
  \label{fig:ea-1_5}}
\end{figure}
\begin{figure}
  \begin{center}
    \begin{picture}(0,0)(-40,100)%
      \includegraphics{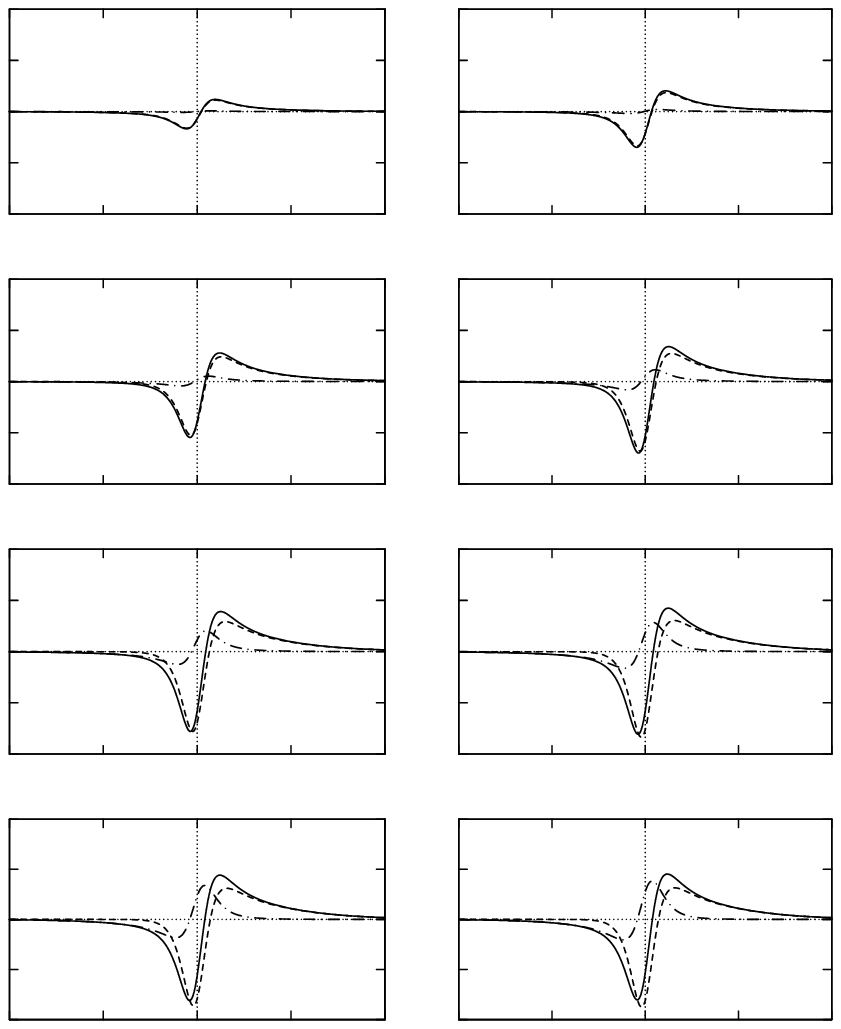}%
    \end{picture}%
    \begingroup
    \setlength{\unitlength}{0.0200bp}%
    \begin{picture}(18000,10800)(-2000,5000)%
      \put(1100,12700){\makebox(0,0)[r]{\strut{}$-$8}}%
      \put(1100,14175){\makebox(0,0)[r]{\strut{}0}}%
      \put(1100,15650){\makebox(0,0)[r]{\strut{}8}}%
      \put(6750,15300){\makebox(0,0)[r]{\strut{}(a)}}%
      \put(1100,8812){\makebox(0,0)[r]{\strut{}$-$8}}%
      \put(1100,10287){\makebox(0,0)[r]{\strut{}0}}%
      \put(1100,11762){\makebox(0,0)[r]{\strut{}8}}%
      \put(6750,11412){\makebox(0,0)[r]{\strut{}(c)}}%
      \put(1100,4924){\makebox(0,0)[r]{\strut{}$-$8}}%
      \put(1100,6399){\makebox(0,0)[r]{\strut{}0}}%
      \put(1100,7874){\makebox(0,0)[r]{\strut{}8}}%
      \put(6750,7524){\makebox(0,0)[r]{\strut{}(e)}}%
      \put(1100,1100){\makebox(0,0)[r]{\strut{}$-$8}}%
      \put(1100,2543){\makebox(0,0)[r]{\strut{}0}}%
      \put(1100,3986){\makebox(0,0)[r]{\strut{}8}}%
      \put(1375,550){\makebox(0,0){\strut{}$-$0.5}}%
      \put(4078,550){\makebox(0,0){\strut{}0.0}}%
      \put(6780,550){\makebox(0,0){\strut{}0.5}}%
      \put(0,8375){\rotatebox{90}{\makebox(0,0)[c]{\strut{}Excitation spectrum, $n_{\sigma D}^{(ex)}$}}}
      \put(4078,0){\makebox(0,0)[c]{\strut{}$\epsilon_D$}}%
      \put(6750,3642){\makebox(0,0)[r]{\strut{}(g)}}%
      \put(13188,15300){\makebox(0,0)[r]{\strut{}(b)}}%
      \put(13188,11412){\makebox(0,0)[r]{\strut{}(d)}}%
      \put(13188,7524){\makebox(0,0)[r]{\strut{}(f)}}%
      \put(7845,550){\makebox(0,0){\strut{}$-$0.5}}%
      \put(10530,550){\makebox(0,0){\strut{}0.0}}%
      \put(13215,550){\makebox(0,0){\strut{}0.5}}%
      \put(10530,0){\makebox(0,0)[c]{\strut{}$\epsilon_D$}}%
      \put(13188,3642){\makebox(0,0)[r]{\strut{}(h)}}%
    \end{picture}%
    \endgroup
  \end{center}
  \vspace{3.5cm}
  \caption{
    As for Fig. \ref{fig:ea-2_5}, but with $\eaD=-0.5$.
  \label{fig:ea-0_5}}
\end{figure}

Figures \ref{fig:ea-2_5}, \ref{fig:ea-1_5} and \ref{fig:ea-0_5} show a series of snapshots of the evolution
of the electronic excitation spectra for each of the model systems.
In each case the majority of the evolution occurs during the period in which the rate of change of $\Gamma_D$ is largest.
The snapshots of the $\eaD=-2.5$ system, in Fig. \ref{fig:ea-2_5}, show the early evolution of $\nsxD$ occurring primarily in the minority
spin-state, while the majority state evolves later.
The $\eaD=-1.5$ model system in contrast, see Fig. \ref{fig:ea-1_5}, has both spin states evolving in a symmetrical manner resulting in
near-identical spectra for majority spin electrons and minority spin holes.
The $\eaD=-0.5$ model system, see Fig. \ref{fig:ea-0_5}, is similar to the $\eaD=-2.5$ system with the
reversal of the spin states; evolution occurs early and late in the majority and minority states respectively.

In each of the model systems the excitation spectra show a number of similar features, with the balance between
them determined by the parameters of the
system. The majority spin spectrum consists of a fairly narrow peak of holes close to the Fermi level with a relatively small tail
extending below $\epsilon_F$. Above the Fermi level, however, the majority spin distribution is flatter and broader, leading to a larger
tail of electronic excitations at higher energies. The overall size of the majority spin distribution is governed by the total change in the
occupation of the relevant adsorbate state. The largest change (from 1.00 to 0.36) occurs for $\eaD=-0.5$ (Fig. \ref{fig:occ}(e)) and
this gives rise to the largest majority-spin excitation spectrum (Fig. \ref{fig:ea-0_5}).
Conversely, the small change in the majority-spin occupation for $\eaD=-2.5$ (from 1.00 to 0.61, Fig. \ref{fig:occ}(a))
yields a relatively small excitation spectrum (Fig. \ref{fig:ea-2_5}).
These trends are reversed for the minority-spin components.
These have a narrow electron peak near the Fermi level with a small number of higher-energy electron excitations.
There is a broader hole distribution, which gives the dominant contribution to higher-energy holes below the Fermi energy.
The magnitude of the minority-spin spectrum is again governed by the total change in the relevant occupation, so that
the largest minority spectrum now occurs for $\eaD=-2.5$.

These results have interesting consequences in the context of chemicurrent generation, where electrons or holes with
sufficient energy to cross a Schottky barrier are detected\cite{nienhaus99}.
Figures \ref{fig:ea-2_5} to \ref{fig:ea-0_5} show that an asymmetry in the adsorbate energy levels with respect to the
Fermi level will lead to an asymmetry in the measured electron and hole currents.
For example, the ratio of electrons to holes at $\epsilon_D=\pm 0.5$ is 1:6.6, 1:1.2 and 5.0:1 for
$\eaD=-$2.5, $-$1.5 and $-$0.5 respectively.
We also note that the high energy tails of $\nsxD$ are dominated by the majority spin for electrons and the minority spin for holes.
For each of the model systems the spectrum consists of at least 96\% majority spin electrons above $\epsilon_D=0.2$,
with similar fraction of minority spin holes below $\epsilon_D=-0.2$.
This implies that a spin-polarised beam of adsorbates made incident on a metal surface would generate a spin-polarised chemicurrent.

\begin{figure}
  \begin{center}
    \begin{picture}(0,0)(40,0)%
      \includegraphics{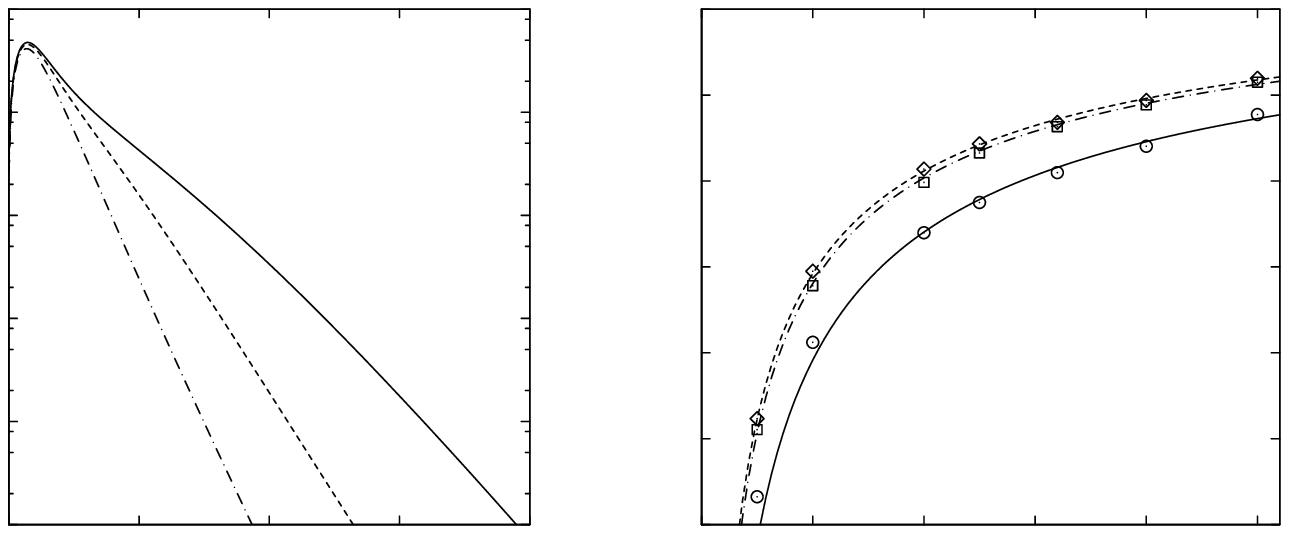}%
    \end{picture}%
    \begingroup
    \setlength{\unitlength}{0.0200bp}%
    \begin{picture}(18000,10800)(2000,0)%
      \put(500,4811){\rotatebox{90}{\makebox(0,0)[c]{\strut{}Excitation spectrum, $n_{\sigma D}^{(ex)}$}}}%
      \put(2200,1100){\makebox(0,0)[r]{\strut{} $10^{-4}$}}
      \put(2200,2584){\makebox(0,0)[r]{\strut{} 0.001}}%
      \put(2200,4069){\makebox(0,0)[r]{\strut{} 0.01}}%
      \put(2200,5553){\makebox(0,0)[r]{\strut{} 0.1}}%
      \put(2200,7038){\makebox(0,0)[r]{\strut{} 1}}%
      \put(2200,8522){\makebox(0,0)[r]{\strut{} 10}}%
      \put(2475,550){\makebox(0,0){\strut{}0.00}}%
      \put(4350,550){\makebox(0,0){\strut{}0.25}}%
      \put(6225,550){\makebox(0,0){\strut{}0.50}}%
      \put(8100,550){\makebox(0,0){\strut{}0.75}}%
      \put(9975,550){\makebox(0,0){\strut{}1.00}}%
      \put(6225,0){\makebox(0,0){\strut{}$\epsilon_D$}}%
      \put(9200,8150){\makebox(0,0)[l]{\strut{}(a)}}%
      \put(10900,4811){\rotatebox{90}{\makebox(0,0)[c]{\strut{}Decay parameter, $\lambda$}}}%
      \put(12175,1100){\makebox(0,0)[r]{\strut{}$-$35}}%
      \put(12175,2337){\makebox(0,0)[r]{\strut{}$-$30}}%
      \put(12175,3574){\makebox(0,0)[r]{\strut{}$-$25}}%
      \put(12175,4811){\makebox(0,0)[r]{\strut{}$-$20}}%
      \put(12175,6048){\makebox(0,0)[r]{\strut{}$-$15}}%
      \put(12175,7285){\makebox(0,0)[r]{\strut{}$-$10}}%
      \put(12175,8522){\makebox(0,0)[r]{\strut{}$-$5}}%
      \put(12450,550){\makebox(0,0){\strut{}0.00}}%
      \put(14051,550){\makebox(0,0){\strut{}0.25}}%
      \put(15652,550){\makebox(0,0){\strut{}0.50}}%
      \put(17253,550){\makebox(0,0){\strut{}0.75}}%
      \put(18854,550){\makebox(0,0){\strut{}1.00}}%
      \put(20455,550){\makebox(0,0){\strut{}1.25}}%
      \put(16445,0){\makebox(0,0)[c]{\strut{}Relative approach speed}}%
      \put(12514,8150){\makebox(0,0)[l]{\strut{}(b)}}%
    \end{picture}%
    \endgroup
  \end{center}
  \vspace{0cm}
  \caption{(a) Electron excitation spectrum at $t_D=50$ for $\epsilon_{aD}=-1.5$ at full speed (solid line), half speed (dashed)
    and quarter speed (dot-dashed line) relative to the calculation in Fig. \ref{fig:ea-1_5}.
    (b) variation of the decay parameter $\lambda$ (see text) with speed for $\epsilon_{aD}=-2.5$ (dashed line and diamonds),
    $\epsilon_{aD}=-1.5$ (solid line and circles) and $\epsilon_{aD}=-0.5$ (dot-dashed line and squares).
    Points are fitted gradients to $\nsxD$ and lines are speed$^{-0.5}$ fits to these data.
  \label{fig:grad}}
\end{figure}

Figure \ref{fig:grad} shows the effect of changing the rate of variation of $\Gamma_D$ on $\nsxD$.
The rate of variation can be interpreted as the approach speed of an adsorbate,
where larger peak gradients of $\Gamma_D$ imply higher speeds.
In Fig. \ref{fig:grad}(a) we have plotted the total electron excitation spectrum for $\eaD=-1.5$ for three different speeds with a
logarithmic scale for $\nsxD$.
For each of the approach speeds the excitation spectrum above $\epsilon_D\approx 0.2$ varies exponentially with energy.
As would be expected, decreasing the speed of approach to the surface reduces the magnitude of $\nsxD$, i.e.
the evolving distribution of occupied states will converge eventually to the instantaneous distribution.
To analyse this further, we have fitted the electronic excitation spectra above $\epsilon_D=0.2$ to the
exponential $e^{\lambda\epsilon_D}$ for a number of different approach speeds.
In Fig. \ref{fig:grad}(b) the decay parameter $\lambda$ is plotted as a function of approach speed for the three model systems.
We find that the variation of the parameter $\lambda$ is well modelled by a speed$^{-0.5}$ dependence for
each of the model systems, with the $\eaD=-1.5$ system having a larger
gradient than the $\eaD=-2.5$ and $-0.5$ systems, which behave similarly.
We have not, to date, been able to explain the origin of this speed$^{-0.5}$ dependence.

\section{Conclusions}
\label{sec:conclusion}

We have presented an analytical solution for the time evoution of the mean-field
Newns-Anderson model, which has enabled us to calculate the spectrum of hot electrons
which are excited in the course of an encounter between an absorbate and a metallic
surface. Although the Newns-Anderson model is a grossly over-simplified description
of any real system, it does have the major advantage that results can be obtained
quickly and that trends can be easily investigated. In this paper we have focussed
on model systems that have a spin transition; in a future publication we will examine
in detail the excitation spectra for H/Cu, H/Ag and O/Ag, all of which exhibit this
transition\cite{prl06}. However, our analysis can be applied
to a wide range of other systems; all that is required is a model for the variation
of the bare energy level and the resonance width. The Newns-Anderson model should
provide a useful semi-quantitative description of the non-adiabatic coupling between
an absorbate and a metallic substrate. This will allow, for example, an investigation
of the validity of a nearly adiabatic (ie friction-based) approach and the
extent to which the forced-oscillator model provides an accurate description of
the excitation spectrum. The Newns-Anderson model will also provide a useful check on 
a more sophisticated theory of electronic excitations, for example, one based on
ab-initio time-dependent density functional theory\cite{lindenblatt06}.

There is one aspect of our work which merits further discussion. One of the key
advantages of using a simple analytical model is that the results obtained from
it are usually transparent, which enables us to obtain physical insights that
can be applied to more complex situations. In our case, however, the expressions
for the time evolution of the electronic distributions become so complicated
that this transparency is lost. Although it is possible to determine where each
term in the final expressions have come from, we have not to date been able to extract
a clear physical picture of the excitation process. Clear trends can be observed
in Figs (2) to (4), and in Fig (5), but going from a description to an explanation
remains a challenge.

\section{Acknowledgements}
\label{sec:acknowledge}

This work has been supported by the UK Engineering and Physical Sciences Research Council.

\appendix
\renewcommand{\thesection}{A}
\renewcommand{\theequation}{A.\arabic{equation}}
\renewcommand{\thesubsection}{A.\arabic{subsection}}
\setcounter{equation}{0}  
\section{Occupation functions}
\label{app:occ_funcs}
Substitution of (\ref{eq:ca_defn}) into (\ref{eq:ck_defn}) yields
\begin{eqnarray}
  \ck(t) &=& -\int^t_{t_0} dt_1 \vak^*(t_1) \exp\left[-i\eks(t-t_1)\right] \int^{t_1}_{t_0} dt_2 \exp\left[-i\int^{t_1}_{t_2} \etilde(t')dt'\right] \nonumber \\
         & &  \hspace{4cm}  \times \sum_{k'} \vakp(t_2) \exp\left[-i\eksp(t_2-t_0)\right]\ckp(t_0) \nonumber \\
         & & -i \int^t_{t_0} dt_1 \vak^*(t_1)\exp[-i\eks(t-t_1)] \exp\left[-i\int^{t_1}_{t_0} \etilde(t')dt'\right] \ca(t_0)\nonumber\\
         & & + \exp[-i\eks(t-t_0)] \ck(t_0).
  \label{eq:ck_defn2}
\end{eqnarray}
The required occupation functions follow by using (\ref{eq:ca_defn}) and (\ref{eq:ck_defn2}) together with the initial conditions
$\langle\!\langle \cad(t_0)\ca(t_0) \rangle\!\rangle = f(\ebar(t_0)) $,
$\langle\!\langle \cad(t_0)\ck(t_0) \rangle\!\rangle = 0 $,
$\langle\!\langle \ckd(t_0)\ckp(t_0) \rangle\!\rangle = f(\epsilon_{k\sigma})
\delta_{kk^\prime}$.
We find
\begin{eqnarray}
  \naks(t)\!\! 
  &\equiv& \langle\!\langle \cad(t)\ck(t) \rangle\!\rangle \nonumber\\
  &=&\!\! -i\!\!\int^t_{t_0} dt_1 \vak^*(t_1) \int d\epsilon' f(\epsilon')\ps^*(\epsilon',t) 
  \ps(\epsilon',t_1)\exp[i(\epsilon'-\eks)(t-t_1)] \nonumber \\
  & &\!\! +i f(\eks)\int^t_{t_0} dt_1 \vak^*(t_1) 
  \exp\!\left[i\int^t_{t_1} (\etilde^*(t')-\eks)dt'\right] \nonumber \\
  & &\!\! -i\nas(t_0)\exp\!\left[i\int^t_{t_0}(\etilde^*(t') - \eks)dt'\right]\nonumber \\
  && \hspace{1cm}\times\int^t_{t_0} dt_1 \vak^*(t_1)
  \exp\!\left[-i\int^{t_1}_{t_0}(\etilde(t') - \eks)dt'\right],
  \label{eq:naks}
\end{eqnarray}
and
\begin{eqnarray}
  \nkkps(t) 
  &\equiv& \langle\!\langle \ckd(t)\ckp(t) \rangle\!\rangle \nonumber\\
  &=&\!\!  \int d\epsilon' f(\epsilon')  \int^t_{t_0}\! dt_1\vak(t_1)\ps^*(\epsilon',t_1)\exp[-i(\eks-\epsilon')(t_1-t)] \nonumber \\
  & & \hspace{1cm}\times\int^t_{t_0}\! dt_2 \vakp^*(t_2)\ps(\epsilon',t_2)\exp[i(\eksp-\epsilon')(t_2-t)]\nonumber \\
  & &\!\! -f(\eksp)\!\!\int^t_{t_0}\! dt_1 \vak(t_1)\exp[i(\eks-\eksp)(t-t_1)]\nonumber \\
  && \hspace{1cm}\times \int^{t_1}_{t_0}\!dt_2\vakp^*(t_2)
  \exp\!\!\left[i\!\!\int^{t_1}_{t_2}(\etilde^*(t')-\eksp)dt'\right]\nonumber \\
  & &\!\! -f(\eks)\!\!\int^t_{t_0}\! dt_1 \vakp^*(t_1)\exp[i(\eks-\eksp)(t-t_1)]\nonumber \\
  && \hspace{1cm}\times\int^{t_1}_{t_0}\!dt_2\vak(t_2)
  \exp\!\!\left[-i\!\!\int^{t_1}_{t_2}(\etilde(t')-\eks)dt'\right]\nonumber \\
  & &\!\! + \nas(t_0)\exp[i(\eks-\eksp)(t-t_0)] \int^t_{t_0} dt_1 \vak(t_1) \exp\left[i\int^{t_1}_{t_0}(\etilde^*(t')-\eks)dt'\right] \nonumber \\
  & & \hspace{3cm} \times\int^t_{t_0} dt_2 \vakp^*(t_2) \exp\left[-i\int^{t_2}_{t_0}(\etilde(t')-\eksp)dt'\right] + f(\eks) \delta_{kk'}. \nonumber \\
  \label{eq:nkkps}
\end{eqnarray}
The remaining occupation function in (\ref{eq:elec_spec_dyn_4}) is $\nkks$, which is a special case of $\nkkps$.
From (\ref{eq:nkkps}) we find
\begin{eqnarray}
  \nkks(t) &=& \int d\epsilon' f(\epsilon') \left| \int^t_{t_0}\! dt_1\vak^*(t_1)\ps(\epsilon',t_1)\exp[i(\eks-\epsilon')(t_1-t)] \right|^2 \nonumber \\
  && - 2f(\eks) \Re\Bigg\{ \int^t_{t_0} dt_1 \vak^*(t_1) \int^{t_1}_{t_0} dt_2 \vak(t_2) \exp\left[-i\int^{t_1}_{t_2}(\etilde(t')-\eks)dt'\right] \Bigg\} \nonumber \\
  && +\nas(t_0) \left| \int^t_{t_0} dt_1 \vak^*(t_1) \exp\left[-i \int^{t_1}_{t_0}(\etilde(t')-\eks)dt'\right] \right|^2 + f(\eks).
  \label{eq:nkks}
\end{eqnarray}

\renewcommand{\thesection}{B}
\renewcommand{\theequation}{B.\arabic{equation}}
\renewcommand{\thesubsection}{B.\arabic{subsection}}
\setcounter{equation}{0}  
\section{Charge conservation of the distribution function $\ns$}
\label{app:charge_cons}
In this appendix we demonstrate that the time-dependent distribution function (\ref{eq:ns_final}) conserves charge by taking
the integral of $\ns$ over all energies.
We will use numerical superscripts to denote the individual terms on the right hand side of (\ref{eq:ns_final}).
The integral over the first term can be written, using (\ref{eq:qs_defn}) and (\ref{eq:Pinst_defn}), as
\begin{eqnarray}
  \!\!\!\!  \int\! d\epsilon \, \nsi(\epsilon,t) &=&\!\! \int d\epsilon' f(\epsilon') \int d\epsilon
  \left| \qs(\epsilon,\epsilon',t) + \Pinst(\epsilon,t)\ps(\epsilon',t)\right|^2 \nonumber \\
  &=&\int\! d\epsilon'f(\epsilon')\!\!\int^t_{t_0}\! dt_1 \int^t_{t_0}\! dt_2 \sqrt{\Gamma(t_1)\Gamma(t_2)} \ps^*(\epsilon',t_2)
  \ps(\epsilon',t_1)\exp[i\epsilon'(t_2-t_1)] \nonumber \\
  && \hspace{1cm}\times\int d\epsilon \frac{\exp[i\epsilon(t_1-t_2)]}{2\pi}\nonumber \\
  & & -2\Re\left\{\gpi{t}\int d\epsilon' f(\epsilon') \ps^*(\epsilon',t)\int^t_{t_0}\!\! dt_1 \gpi{t_1}\ps(\epsilon',t_1)
  \exp[i\epsilon'(t-t_1)] \right.\nonumber \\
  && \hspace{1cm}\times\left.\int d\epsilon \frac{\exp[-i\epsilon(t-t_1)]}{\epsilon -\etilde^*(t)}\right\} \nonumber \\
  & & + \int d\epsilon' f(\epsilon') \left|\ps(\epsilon',t)\right|^2 \int d\epsilon \frac{\Gamma(t)}{2\pi[(\epsilon-\ebar(t))^2+\Gamma(t)^2/4]}.
  \label{eq:chg_cons_1a}
\end{eqnarray}
The $\epsilon$ integral in the first term of this expression gives the delta function $\delta(t_1-t_2)$ and that in the third
is unity. The $\epsilon$ integral in the second term can be evaluated using contour methods, with the contour closed in the lower half plane.
By the residue theorem this integral is zero, because the pole at $\epsilon=\etilde^*(t)=\ebar(t)+i\Gamma(t)/2$ is in the upper half plane.
Equation (\ref{eq:chg_cons_1a}) therefore simplifies to
\begin{eqnarray}
  \int d\epsilon \, \nsi(\epsilon,t) = \int_{t_0}^t dt_1 \Gamma(t_1) \int d\epsilon' f(\epsilon') \left|\ps(\epsilon',t_1)\right|^2
  + \int d\epsilon' f(\epsilon') \left| \ps(\epsilon',t)\right|^2.
  \label{eq:chg_cons_1}
\end{eqnarray}
The integral of the second term in (\ref{eq:ns_final}) results in a term which we cannot simplify at this point so we simply state
\begin{eqnarray}
  \int d\epsilon\, \nsii(\epsilon,t) = -2 \int^t_{t_0} dt_1 \gpi{t_1}\int d\epsilon f(\epsilon)\Re\{ \ps(\epsilon,t_1)\}.
  \label{eq:chg_cons_2}
\end{eqnarray}
The third term in (\ref{eq:ns_final}) can be integrated to give
\begin{eqnarray}
  \int d\epsilon \, \nsiii(\epsilon,t) &=& \nas(t_0) \int d\epsilon \left\vert\rs(\epsilon,t) + \Pinst(\epsilon,t)
  \exp\left[-\frac{1}{2}\int^t_{t_0}\Gamma(t')dt'\right]\right\vert^2 \nonumber \\
  & = & \nas(t_0) \int d\epsilon |\rs(\epsilon,t)|^2 \nonumber \\
  && + 2\nas(t_0)\exp\left[-\frac{1}{2}\int^t_{t_0}\Gamma(t')dt'\right] \gpi{t}\Im\left\{
  \int d\epsilon \frac{\rs^*(\epsilon,t)}{\epsilon-\etilde(t)}\right\} \nonumber \\
  & &+ \nas(t_0)\exp\left[-\int^t_{t_0} \Gamma(t')dt'\right]\int d\epsilon \frac{\Gamma(t)}{2\pi[(\epsilon-\ebar(t))^2+\Gamma(t)^2/4]}.
  \label{eq:chg_cons_3a}
\end{eqnarray}
Using contour methods, and the definition of $\rs$ (\ref{eq:rs_defn}), this expression can be evaluated, yielding
\begin{equation}
  \int d\epsilon \, \nsiii(\epsilon,t) = \nas(t_0) \int^t_{t_0} dt_1 \Gamma(t_1) \exp\!\!\left[-\!\int^{t_1}_{t_0}\Gamma(t')dt'\right] + \nas(t_0) \exp\!\!\left[-\!\int^{t}_{t_0}\Gamma(t')dt'\right].
  \label{eq:chg_cons_3}
\end{equation}
The fourth term in (\ref{eq:ns_final}) will give us the number of metal electrons, $\Nk$:
\begin{equation}
  \int d\epsilon \, \nsiv(\epsilon,t) = \int d\epsilon \sum_k f(\eks) \delta(\epsilon-\eks) = \sum_k f(\eks) = \Nk.
  \label{eq:chg_cons_4}
\end{equation}
The fifth, sixth and seventh terms in (\ref{eq:ns_final}) can easily be shown to integrate to zero using contour methods.

By combining (\ref{eq:chg_cons_1}), (\ref{eq:chg_cons_2}), (\ref{eq:chg_cons_3}) and
(\ref{eq:chg_cons_4}), and using the definition of $\nas(t)$
in (\ref{eq:nas_defn}), we obtain the following expression for the total charge;
\begin{eqnarray}
  \int d\epsilon \, \ns(\epsilon,t) &=&  \int^t_{t_0} dt_1 \Gamma(t_1) \int d\epsilon' f(\epsilon') \left|\ps(\epsilon',t_1)\right|^2
  + \int d\epsilon' f(\epsilon') \left| \ps(\epsilon',t)\right|^2 \nonumber \\
  & &-2 \int^t_{t_0} dt_1 \gpi{t_1}\int d\epsilon f(\epsilon) \Re\{\ps(\epsilon,t_1)\}\nonumber \\
  &&+ \int^t_{t_0} dt_1 \Gamma(t_1) \nas(t_0)\exp\!\!\left[-\!\int^{t_1}_{t_0}\Gamma(t')dt'\right]\nonumber \\
  & & + \nas(t_0) \exp\!\!\left[-\!\int^{t}_{t_0}\Gamma(t')dt'\right]  + \Nk\nonumber\\
  & = & \int^t_{t_0} dt_1 \Gamma(t_1) \nas(t_1) + \nas(t) -2 \int^t_{t_0} dt_1 \gpi{t_1}\int d\epsilon f(\epsilon) \Re\left\{\ps(\epsilon,t_1)\right\}\nonumber \\
  &&+\Nk.
  \label{eq:chg_cons_Fa}
\end{eqnarray}
To remove the time integrals from (\ref{eq:chg_cons_Fa}) we consider the time derivative of $\nas(t)$.
Using (\ref{eq:nas_defn}) this becomes
\begin{eqnarray}
  \frac{d}{dt}\nas(t)\!\! &=&\!\! \nas(t_0)\frac{d}{dt}\exp\left[-\int^t_{t_0} \Gamma(t')dt'\right] + \int d\epsilon f(\epsilon) \frac{d}{dt} \vert \ps(\epsilon,t)\vert^2 \nonumber \\
    &=& \!\! -\Gamma(t)\nas(t) + 2\gpi{t}\int d\epsilon f(\epsilon) \Re\lbrace\ps(\epsilon,t)\rbrace.
  \label{eq:chg_cons_Fb}
\end{eqnarray}
By integrating over time from $t_0$ to $t$ this yields
\begin{eqnarray}
  \nas(t)-\nas(t_0) = -\int^t_{t_0} dt_1 \Gamma(t_1) \nas(t_1) + 2 \int^t_{t_0} dt_1 \gpi{t_1}\int d\epsilon f(\epsilon) \Re\{\ps(\epsilon,t_1)\},
  \label{eq:chg_cons_Fc}
\end{eqnarray}
which, on combination with (\ref{eq:chg_cons_Fa}), gives
\begin{equation}
  \int d\epsilon \, \ns(\epsilon,t) = \nas(t_0) + \Nk.
  \label{eq:chg_cons_final}
\end{equation}
This confirms that charge is conserved, i.e. the number of electrons of spin $\sigma$ is the sum of those in the adsorbate and the metal
prior to any interaction and does not vary with time.

\end{spacing}
\end{document}